%% file: main.tex
\documentclass{aastex63}

\usepackage{amsmath}
\usepackage{braket}
\usepackage{bm}
\usepackage{amsmath,amssymb}
\usepackage{subfigure}
\usepackage{appendix}
\usepackage{capt-of}
\usepackage{booktabs}
\usepackage{varwidth}
\usepackage{graphicx}
\usepackage{paralist}
\usepackage{xcolor}
\usepackage[normalem]{ulem}

\newcommand{\mode}[2]{\ensuremath{_{#1}\mathrm{S}_{#2}}\,}

\newcommand{\cL}{\mathcal{L}\,}
\newcommand{\stackm}{\mathcal{S}_m}

\newcommand{\grad}{\ensuremath{\bm{\nabla}}}

\newcommand{\bnabla}{\ensuremath{\bm{\nabla}}}

\newcommand{\xiv}{\boldsymbol{\xi}}

\newcommand{\ev}[1]{\hat{\bm{e}}_{#1}}

\newcommand{\solintv}{\int_{\odot} \mathrm{d}^3\mathbf{r}\,}

\input{newcommands}

\submitjournal{\apjs}
\received{December 23, 2021} 
\revised{January 18, 2021}
\accepted{January 21, 2021}

\shorttitle{Bayesian inference of differential rotation}
\shortauthors{Kashyap and Bharati Das et al.}

\begin{document}
\title{Inferring solar differential rotation through normal-mode coupling using Bayesian statistics}

\correspondingauthor{Samarth G. Kashyap}
\email{g.samarth@tifr.res.in}

\thanks{Both authors have contributed equally to this study.}
\author[0000-0001-5443-5729]{${}^*$Samarth G. Kashyap}
\affil{Department of Astronomy and Astrophysics \\
Tata Institute of Fundamental Research \\
Mumbai, India}

\author[0000-0003-0896-7972]{${}^*$Srijan Bharati Das}
\affiliation{Department of Geosciences \\
Princeton University \\
Princeton, New Jersey, USA}

\author[0000-0003-2896-1471]{Shravan M. Hanasoge}
\affil{Department of Astronomy and Astrophysics \\
Tata Institute of Fundamental Research \\
Mumbai, India}

\author[0000-0002-3710-7245]{Martin F. Woodard}
\affiliation{NorthWest Research Associates \\
Boulder Office, 3380 Mitchell Lane \\
Boulder, Colorado, USA}

\author[0000-0002-2742-8299]{Jeroen Tromp}
\affiliation{Department of Geosciences \\
and Program in Applied \& Computational Mathematics \\
Princeton University \\
Princeton, New Jersey, USA}



\begin{abstract}
Normal-mode helioseismic data analysis uses observed solar oscillation spectra 
to infer perturbations in the solar interior due to global and 
local-scale flows and structural asphericity. Differential rotation, 
the dominant global-scale axisymmetric perturbation, has been tightly 
constrained primarily using measurements of frequency splittings via 
``$a$-coefficients". However, the frequency-splitting formalism invokes the approximation that multiplets are isolated. This assumption is inaccurate for modes at high angular degrees. Analysing eigenfunction corrections, which respect cross coupling of modes across multiplets, is a more accurate approach. However, applying standard inversion techniques using these cross-spectral 
measurements yields $a$-coefficients with a significantly wider spread than the well-constrained results from frequency splittings. 
In this study, we apply Bayesian statistics to infer 
$a$-coefficients due to differential rotation 
from cross spectra for both $f$-modes and $p$-modes.
We demonstrate that this technique works reasonably well for modes
with angular degrees~$\ell=50-291$. The inferred
$a_3-$coefficients are found to be within $1$~nHz of the
frequency splitting values for~$\ell > 200$. We also show that the technique fails at~$\ell < 50$ owing
to the insensitivity of the measurement to the perturbation.
These results serve to further establish mode coupling as an 
important helioseismic technique with which to infer internal 
structure and dynamics, both axisymmetric 
(e.g., meridional circulation) and non-axisymmetric perturbations. 
\end{abstract}

\keywords{Sun: helioseismology --- Sun: oscillations --- 
Sun: interior --- differential rotation --- MCMC}


\section{Introduction} \label{sec:intro}

The strength and variation of observed solar activity is governed by 
the spatio-temporal dependence of flow fields in the convective 
envelope \citep{charbonneau05,fan09}. 
Thus, understanding the physics that governs the evolution 
and sustenance of the activity cycle of the Sun necessitates 
imaging its internal layers. 
While differential rotation has the most significant 
imprint on Dopplergram images \citep{schou98}, 
signatures due to weaker effects, such as meridional circulation
\citep{giles97,BasuAntia99,zhao04,gizon2020_sci} and 
magnetic fields \citep{gough90,goode04,AntiaChitre13},
are also noticeable. 
The ability to image these weaker effects therefore critically 
depends on an accurate measurement of the dominant flows. 
This makes inferring the strength of the dominant flows 
along with assigning appropriate statistical uncertainties 
an important area of study.

Differences between normal modes of the Sun and those predicted using standard solar models may be used to constrain solar internal properties.
 The standard models are typically adiabatic, hydrodynamic, 
spherically symmetric and non-rotating, also referred to as
 SNRNMAIS \citep{lavely92,jcd}.  The usual labelling convention, 
 using 3 quantum numbers, $(n,\ell,m)$, where~$n$ denotes radial order,~$\ell$ the angular degree, and~$m$ the azimuthal order, are used to uniquely identify normal modes.
 Departures of solar structure from the SNRNMAIS are
 modelled as small perturbations \citep{jcd_notes_orig}, which ultimately manifest themselves
 as observable shifts (or splittings) in the eigenfrequencies and
 distortions in the eigenfunctions \citep{woodard89}. The distorted
 eigenfunctions may be expressed as a linear combination of reference
 eigenfunctions and are said to be coupled with respect to the reference.
Observed cross-spectra of spherical-harmonic time
series corresponding to full-disk Dopplergrams are used to measure 
eigenfunction distortion. In the present study, we use observational
data from the \emph{Helioseismic Magnetic Imager} (HMI) onboard the \emph{Solar Dynamics Observatory} \citep{schou-hmi-2012}.

Different latitudes of the Sun rotate at different angular 
velocities, with the equator rotating faster than the poles
\citep{howard84,ulrich1988}.
To an observer in a frame co-rotating at a specific rotation 
rate $\bar{\Omega}$ of the Sun, this latitudinal rotational 
shear is the most significant perturbation to the reference model. 
This large-scale toroidal flow $\Omega\,(r,\theta)$ is well approximated 
as being time-independent \citep[shown to vary less than 5\% over the 
last century in][]{gilman74,howard84,basu_antia_2003} and zonal, 
with variations only along the radius $r$ and co-latitude $\theta$. 

Very low $\ell \leq 5$ modes penetrate the deepest layers of the Sun and were used in earlier attempts to constrain the rotation rate in 
the core and radiative interior \citep{claverie81,chaplin99,eff-darwich02,
couvidat03,chaplin04}.
However, observed solar activity is believed to be governed 
by the coupling of differential rotation and magnetic fields 
in the bulk of the convection zone \citep{miesch05}. 
Subsequently, studies using intermediate~$\ell \leq 100$ 
\citep{duvall84,brown_morrow87,brown89,libbrecht89,Duvall1996} and modes 
with relatively high $\ell \leq 250$ \citep{thompson96,kosovichev97_mdi,
schou98} yielded overall convergent results for the rotation profile.
Among other features of the convection zone \citep{howe09}, these 
studies established the presence of shear layers at the base of the
convection zone (the tachocline) and below the solar surface. 

Most of these studies used measurements of frequency splittings in a
condensed convention known as $a$-coefficients \citep{ritzwoller}. 
The azimuthal and temporal independence make differential rotation 
particularly amenable to inversion via $a$-coefficients. 
The assumption behind this formalism is that multiplets, identified by ~$(n,\ell)$, are well separated in frequency from 
each other, known as the `isolated multiplet approximation'. 
This assumption holds true when differential rotation is the 
sole perturbation under consideration \citep{lavely92}, 
even at considerably high~$\ell$. 
 We therefore state at the outset that estimates of $a$-coefficients
determined from frequency splitting serve as reliable measures of
differential rotation \citep{chatterjee-antia-2009}.
Nevertheless, the estimation of non-axisymmetric perturbations
requires a rigorous treatment honoring the cross coupling of multiplets \citep{hanasoge17_etal,sbdas20}. 
In such cases, measuring changes to the eigenfunctions is far more 
effective than, for instance, the $a$-coefficient formalism. 
As a first step, it is therefore important to explore the
potential of eigenfunction corrections to infer differential rotation. 

The theoretical modeling of eigenfunction corrections for given
axisymmetric -- zonal and meridional -- flow fields may be traced back to
\cite{woodard89}, followed up by further investigations
\cite{Woodard00,Gough10,vorontsov07,schad11}. 
\cite{schad13} and \cite{schad20} used
observables in the form of mode-amplitude ratios to infer meridional
circulation and differential rotation, respectively. In this study, we
adopt the closed-form analytical expression for correction coefficients
first proposed by \cite{vorontsov07} and subsequently 
verified to be accurate up to angular degrees as high as~$\ell \leq 1000$
\cite{vorontsov11}, henceforth V11. The method of using cross-spectral
signals to fit eigenfunction corrections was first applied by
\cite{woodard13}, henceforth W13, to infer differential rotation and
meridional circulation. A simple least-squares fitting, assuming a unit
covariance matrix, was used for inversions in W13. Their results of odd
$a$-coefficients (which encodes differential rotation), even though
qualitatively similar, show a considerably larger spread than the results
from frequency splittings. Moreover, the authors of W13 note that the
inferred meridional flow was ``less satisfactory" [than their zonal flow
estimates]. Cross spectra are dominated by differential rotation, 
a much larger perturbation than meridional circulation. Although zonal
and meridional flows are measured in different cross-spectral channels, 
the inference of meridional flow is affected by differential rotation
through leakage. Thus,
the  accurate determination of odd $a$-coefficients is critical to 
the inference of meridional flow.
The relatively large spread in inferences of differential rotation 
obtained by W13 may be due to 
(a) a poorly conditioned minimizing function with multiple local 
minima surrounding the expected (frequency splitting) minima, 
(b) a relative insensitivity of various modes to differential rotation, resulting in a flat minimizing function close to the expected minima, 
(c) an inaccurate estimation of the minimizing function on account 
of assuming a unit-data covariance matrix, and/or 
(d) eigenfunction corrections only yielding accurate results 
in the limit of large $\ell$ ($>250$), where the isolated-multiplet
approximation starts worsening.

In this study, we investigate the above issues and explore the 
potential of using eigenfunction corrections as a means to infer 
differential rotation using tools from Bayesian statistics. 
We apply the Markov Chain Monte Carlo (MCMC) algorithm
\citep{metropolis-ulam-1949,metropolis-etal-1953} using a
minimizing function calculated in the L2 norm, adequately weighted by data
variance. We do not bias the MCMC sampler in light of any previous
measurement, effectively using an uninformed prior. The results inferred,
therefore, are an independent measurement constrained only by observed cross
spectra. Since Bayesian inference is a probabilistic approach to parameter 
estimation, we obtain joint probability-density functions in 
the $a$-coefficient space. This allows us to rigorously compute 
uncertainties associated with the measurements.
We compare and qualify the results obtained with independent 
measurements from frequency splitting and those obtained using similar
cross-spectral analysis in W13.
Further, we report the inadequacy of this method for 
low angular-degree modes on account of poor sensitivity of 
spectra to rotation via $a$-coefficients. 

The structure of this paper is as follows. We establish mathematical
notations and describe the basic physics of normal-mode helioseismology in
Section~\ref{sec:basic_framework}. The governing equations which we use for modeling cross spectra using eigenfunction-correction
coefficients are outlined in Section~\ref{sec:vorontsov_theory}.
Section~\ref{sec:data_analysis} elaborates the steps for computing 
the observed cross spectra and building the misfit function and
estimating data variance for performing the MCMC. Results are discussed
in Section~\ref{sec:results}. Using the $a$-coefficients inferred
from MCMC, cross-spectra are reconstructed in 
Section~\ref{sec:reconst-spectra}. A discussion on sensitivity of 
the current model to the model parameters is presented in 
Section~\ref{sec:a-coeff-sens}. The conclusions from this work are
reported in Section~\ref{sec:conclusion}.

\section{Theoretical Formulation} \label{sec:theory}

\subsection{Basic Framework and Notation} \label{sec:basic_framework}

For inferring flow profiles in the solar interior, we begin by considering the system of coupled hydrodynamic equations, namely,
\begin{eqnarray}
    \partial_t \rho &=& - \bnabla \cdot (\rho\, \mathbf{v}), \label{eqn: HD1} \\
    \rho (\partial_t \mathbf{v} + \mathbf{v}\cdot \bnabla \mathbf{v} ) &=& - \bnabla P - \rho \bnabla \phi, \label{eqn: HD2}\\
    \partial_t P &=& - \mathbf{v}\cdot \bnabla P - \gamma\, P\, \bnabla \cdot \mathbf{v} \label{eqn: HD3}, 
\end{eqnarray}
where $\rho$ is the mass density, $\mathbf{v}$ the material velocity, $P$
the pressure, $\phi$ the gravitational potential and $\gamma$ the ratio
of specific heats determined by an adiabatic equation of state. The
eigenstates of the Sun are modeled as linear combinations of the eigenstates of a standard solar model. Here we use model S as this reference state, which is discussed in \cite{jcd}. In absence of background flows, $\tilde{\bfv} = {\bf 0}$, the
zeroth-order hydrodynamic equations trivially reduce to the hydrostatic equilibrium
$\bnabla \tilde{P} + \tilde{\rho} \bnabla \tilde{\phi} = 0$. Hereafter, all zeroth-order
static fields, unperturbed mode eigenfrequencies, eigenfunctions and amplitudes
corresponding to the reference model will be indicated using tilde
(to maintain consistency with notation used in W13).
In response to small perturbations to the static reference model, 
the system exhibits oscillations $\bfxi (\mathbf{r},t)$. 
These oscillations may be decomposed into resonant ``normal modes" 
of the system, labeled by index $k$, with characteristic
frequency $\tilde\omega_k$ and spatial pattern $\tilde\bfxi_k$, as follows:
\begin{equation}
    \xiv (\boldsymbol{r},t) = \sum_k 
    \tilde\Lambda_k(t)\,\tilde\xiv_k(\boldsymbol{r})\exp(i\tilde\omega_k t),
\end{equation} 
where $\tilde\Lambda_k$ are the respective mode amplitudes
and $\boldsymbol{r} = (r,\theta,\phi)$ denote spherical-polar coordinates.
Linearizing eqns.~(\ref{eqn: HD1})--(\ref{eqn: HD3}) about the hydrostatic background model gives \citep[for a detailed derivation refer
to][]{jcd_notes_orig} 
\begin{equation} \label{eqn:sol_wave_eqn}
    \mathcal{L}_0 \tilde{\xiv}_{k}  = 
    - \grad (\tilde{\rho} c_s^2\, \bnabla \cdot \tilde{\boldsymbol{\xi}}_{k} -
    \tilde{\rho} g\, \tilde{\boldsymbol{\xi}}_{k} \cdot \ev{r}) - g \,\ev{r}
    \bnabla \cdot (\tilde{\rho}\, \tilde{\boldsymbol{\xi}}_{k}) 
    = \tilde{\rho} \,\tilde{\omega}_{k}^2\, \tilde{\xiv_{k}}.
\end{equation}
Here $\tilde{\rho}(r), c_s(r)$, and $g(r)$ denote density, sound
speed, and gravity (directed radially inward) respectively of the
reference solar model, and $\cL_0$ is the self-adjoint unperturbed 
wave operator. This ensures that the eigenfrequencies $\tilde{\omega}_{k}$ 
are real and eigenfunctions $\tilde{\bfxi}_{k}$ are orthogonal. 
Introducing flows and other structure
perturbations through the operator $\delta \cL$ (e.g., magnetic fields or ellipticity) modifies the unperturbed wave equation~(\ref{eqn:sol_wave_eqn}) to
\begin{equation} \label{eqn:total_wave_op}
    \tilde{\rho} \,\omega_k^2\, \xiv_k = \left(\cL_0 + \delta \cL \right) \xiv_k,
\end{equation}
where $\omega_k = \tilde{\omega}_{k} + \delta \omega_k$ and 
$\bfxi_k = \sum_{k'} c_{k'} \tilde{\bfxi}_{k'}$ are the eigenfrequency 
and eigenfunction associated with the perturbed wave
operator $\cL_0 + \delta \cL$. 
The Sun, a predominantly hydrodynamic system, is thus treated as a fluid body with vanishing shear modulus \citep{DT98}. 
This is unfavourable for sustaining shear waves and therefore 
the eigenfunctions of the reference model are very well approximated 
as spheroidal \citep{kendall},
\begin{eqnarray} 
\tilde{\xiv}_k(r,\theta,\phi) = {}_nU{}_{\ell}(r) \,Y_{\ell m}
(\theta,\phi) \,\ev{r} + {}_nV{}_{\ell}(r) \, 
\grad_1 Y_{\ell m}(\theta,\phi). \label{eqn: xi_exp}
 \end{eqnarray}
$\boldsymbol{\grad}_1 = \ev{\theta}\,\partial_{\theta} +
\ev{\phi}\,(\sin\theta)^{-1}\partial_{\phi}$ is the dimensionless lateral covariant derivative operator. Suitably normalized
eigenfunctions $\tilde{\bfxi}_k$ and $\tilde{\bfxi}_{k'}$, where 
$k' = (n',\ell',m')$, satisfy the orthonormality condition
\begin{equation} \label{eqn: orthonormality}
  \solintv \rho\,\boldsymbol{\tilde{\xi}}_{k'}^* \cdot
  \boldsymbol{\tilde{\xi}}_{k} = \delta_{n'n}\, 
  \delta_{\ell' \ell}\, \delta_{m' m}.
\end{equation}
Since we observe only half the solar surface, orthogonality cannot be used to extract each mode
separately. Windowing in the spatial domain results in spectral
broadening, where contributions from neighbouring modes seep into the 
observed mode signal $\varphi^{\ell m}(\omega)$, as described by the 
leakage matrix \citep{schou94},
\begin{equation} \label{eqn: leakage}
    \varphi^{\ell m}(\omega) = \sum_{k'} L^{\ell m}_{k'} \, 
    \Lambda^{k'}(\omega) = \sum_{k'} \tilde{L}^{\ell m}_{k'} \,\
    \tilde{\Lambda}^{k'}(\omega).
\end{equation}
Here, leakage matrices $L^{\ell m}_{k'}, \tilde{L}^{\ell m}_{k'}$ and amplitudes $\Lambda^{k'}(\omega), \tilde{\Lambda}^{k'}(\omega)$ 
of the observed
surface velocity field $\bfv(\omega)$  correspond to the bases of
perturbed ($\bfxi_{k'}$) and unperturbed eigenfunctions
($\tilde{\bfxi}_{k'}$), respectively,
\begin{equation}
    \bfv = \sum_{k'} \Lambda^{k'} \bfxi_{k'} = \sum_{k'}
    \tilde{\Lambda}^{k'} \tilde{\bfxi}_{k'}.
\end{equation}

Since leakage falls rapidly with increasing spectral distance 
$(|\ell - \ell'|, |m-m'|)$, Eqn.~(\ref{eqn: leakage}) demonstrates 
the entangling of modes in spectral proximity to $(\ell,m)$. 
The presence of a zeroth-order flow field $\tilde{\bfv}$ in
Eqns.~(\ref{eqn: HD1})--(\ref{eqn: HD3}) gives rise to perturbed
eigenfunctions $\bfxi_{k}$ and therefore introduces correction factors
$c_{k}^{k'}$ with respect to the unperturbed eigenfunctions
$\tilde{\bfxi}_{k'}$.
\begin{equation} \label{eqn: eigfn_corr}
    \bfxi_k = \sum_{k'} c_k^{k'} \tilde{\bfxi}_{k'}.
\end{equation}

\begin{figure}[!ht]
    \centering
    \includegraphics[width=\textwidth]{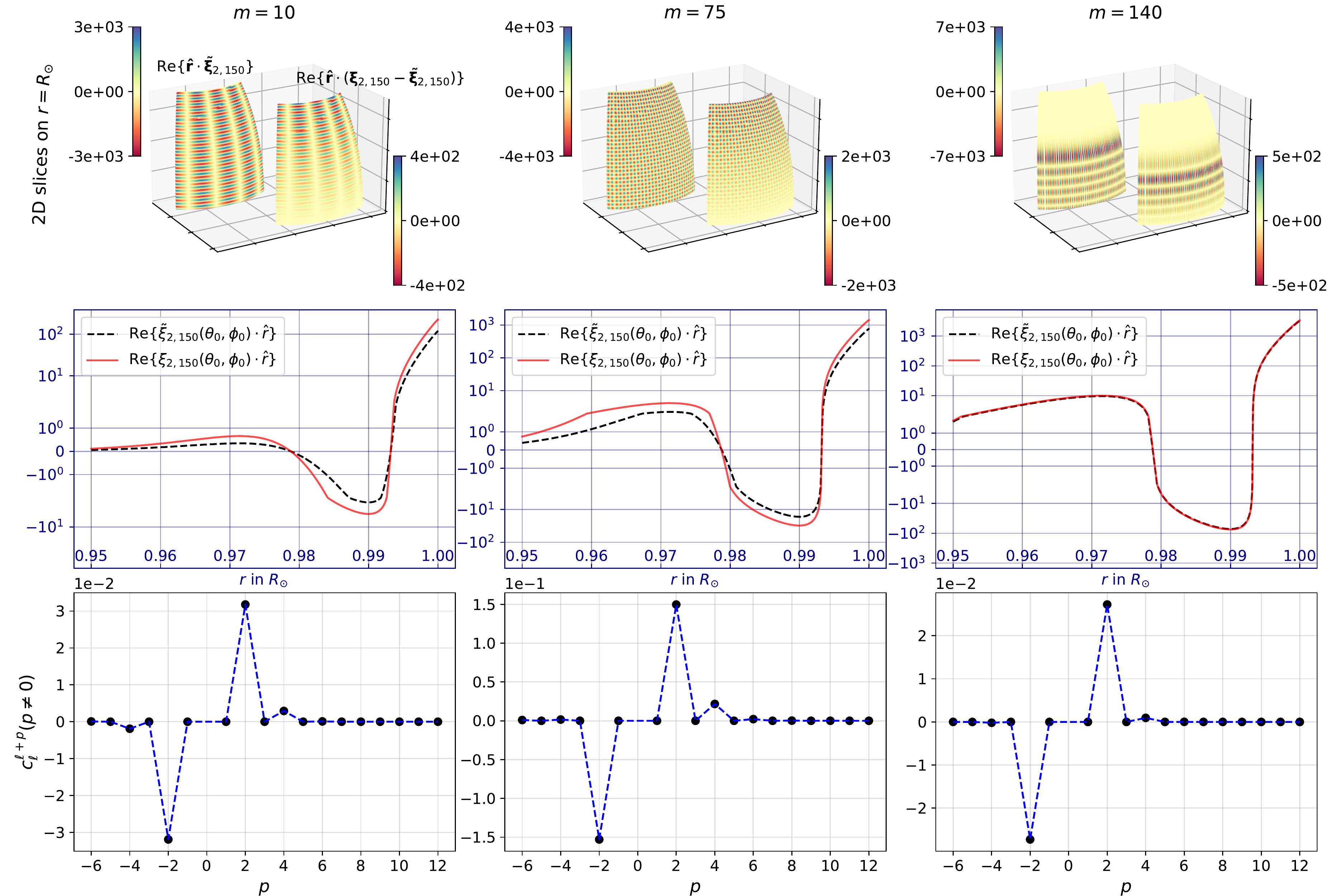}
    \caption{Differential rotation induces 3D distortions in radial eigenfunction of unperturbed mode $(n, l)=(2, 150)$ for 
    $m=10, 75, 140$ at radii $r/R_\odot = 0.95, 1.0$.
    Each column in the \textit{upper} panel correspond to 2D surfaces
    for the undistorted eigenfunctions $\tilde{\vec{\xi}}_{nlm}$ in the 
    \textit{left} slice
    and differences between distorted and undistorted eigenfunctions 
    $\hat{r} \cdot (\vec{\xi}_{nlm} - \tilde{\vec{\xi}}_{nlm})$ in the 
    \textit{right} slice.
    The \textit{middle} panel shows the difference in the radial 
    variation of eigenfunctions for a chosen 
    $(\theta_0,\phi_0) = (67.8^{\circ},177.6^{\circ})$. 
    The \textit{lower} panels indicate
    the magnitudes of the coupling coefficients that induce 
    eigenfunction distortion, as in Eqn.~(\ref{eqn: eigfn_corr}). 
    The self-coupling coefficients $c^{\ell m}_{\ell m}$ 
    (i.e., $p=0$), being the most dominant, are not shown, in order to 
    highlight the contributions of cross-coupling coefficients 
    ($p\neq 0$.)} 
    \label{fig:efn_pert}
\end{figure}

Using this, the statistical expectation of the cross-spectral measurement is expressed as in Eqns.~(14)--(17) of W13,
\begin{equation} \label{eqn: mode_coupling}
    \langle \varphi^{\ell' m'} \varphi^{\ell m} \rangle = 
    \sum_{i,j,k} \tilde{L}^{\ell'm'}_{j} \, 
    \tilde{L}^{\ell m *}_{k} \, c^j_i \, c^{k*}_i \, 
    \langle |\Lambda^i(\omega)|^2 \rangle,
\end{equation}
where $ \langle |\Lambda^i(\omega)|^2 \rangle$ denotes Lorentzians centered at resonant frequencies $\omega = \omega_i$ corresponding to the perturbed modes $\bfxi_{i}$.

\subsection{Eigenfunction corrections due to axisymmetric flows} \label{sec:vorontsov_theory}

This study uses the fact that eigenfunction-correction factors
$c_k^{k'}$ in Eqn.~(\ref{eqn: eigfn_corr}) carry information about the flow field $\tilde{\bfv}$.  Although this problem was first addressed by
\cite{woodard89}, a rigorous treatment using perturbative analysis of mode
coupling was only presented in V11. In this section, we outline the governing
equations for the eigenfunction-correction factors $c_k^{k'}$ due to
differential rotation and meridional circulation as shown in V11. 
Upon introducing flows, the model-S eigenfunctions are corrected as follows:
\begin{equation}
   \label{eqn:voront-pert}
    \bfxi_{\ell} = \sum_{\ell'} c_{\ell}^{\ell'} \, \tilde{\bfxi}_{\ell'} + \delta \bfxi_\ell = \sum_{p = 0,\pm 1, \pm 2, ...} c_{\ell}^{\ell+p} \, \tilde{\bfxi}_{\ell+p} + \delta \bfxi_\ell,
\end{equation}
where $p = \ell' - \ell$ is used to label the offset (in angular degrees)
of the neighbouring mode contributing to the distortion of the erstwhile unperturbed eigenfunction
$\bfxi_{\ell}$ --- visual illustration may be found in Figure~\ref{fig:efn_pert}. Correction factors $c_{\ell}^{\ell+p}$ solely from modes 
with the same radial orders and azimuthal degrees are considered in
Eqn.~(\ref{eqn:voront-pert}) and therefore labels $n$ and $m$ are suppressed.
$c_{\ell,m}^{\ell+p,m'} = 0$ for $m \neq m'$ since differential rotation and meridional circulation are
axisymmetric (see selection rules imposed due to Wigner 3-$j$ symbols in
Appendix~A of V11). Corrections from modes belonging to a different radial order $n$ are
accumulated in $\delta \bfxi$. Following V11 and W13, subsequent treatment
ignores terms in $\delta \bfxi$ since it is considered to be of the order of the
perturbation $\delta \cL$ or smaller (rendering them at least second order in perturbed 
quantities). This is because the correction factor
$c_{n \ell}^{n' \ell'}$ is non-trivial only if modes $\mode{n}{\ell}$ and
$\mode{n'}{\ell'}$ are proximal in frequency space as well as the angular
degree $s$ of the perturbing flow satisfies the relation 
$|\ell' - \ell| \leq s$. 
For modes belonging to different dispersion branches $(n \neq n')$, with
either $\ell$ or $\ell'$ being moderately large ($> 50$) the prior conditions
are not satisfied, since, for differential rotation, the largest
non-negligible angular degree of perturbation is $s = 5$. 

As shown in V11, using eigenfunction perturbations as in
Eqn.~(\ref{eqn:voront-pert}) and eigenfrequency perturbations 
$\omega_{\ell} = \tilde{\omega}_{\ell} + \delta \omega_{\ell}$, the wave
equation~(\ref{eqn:total_wave_op}) reduces to an eigenvalue problem 
of the form
\begin{equation}
    \mathbf{Z}\, \boldsymbol{\mathcal{C}}_{\ell} = \delta \omega_{\ell} \, \boldsymbol{\mathcal{C}}_{\ell},
\end{equation}
where 
$\boldsymbol{\mathcal{C}}_{\ell} = \{...,c_{\ell}^{\ell-1},c_{\ell}^{\ell},
c_{\ell}^{\ell+1},...\}$ are eigenvectors corresponding to the 
$(P \times P)$ self-adjoint matrix $\bfZ$ and $P = \mathrm{max}(|\ell'-\ell|)$ 
denotes the largest offset of a contributing mode $\ell'$ from $\ell$ according as Eqn.~(\ref{eqn:voront-pert}). From detailed considerations of first- and second-order quasi-degenerate perturbation theory, V11 showed that the following closed-form expression for correction coefficients is accurate up to angular degrees as high as $\ell = 1000$:
\begin{equation}
    c_{\ell}^{\ell+p} = \tfrac{1}{\pi} \int_0^{\pi} \cos \left[pt - \sum_{k=1,2,...}\tfrac{2}{k}\mathrm{Re}(b_k) \sin{(kt)} \right] \times \mathrm{exp}\left[i \sum_{k=1,2,...} \tfrac{2}{k} \mathrm{Im}(b_k) \cos{(kt)} \right] \mathrm{d}t, \qquad p=0,\pm1,...
    \label{eqn:clp},
\end{equation}
where the convenient expressions for real and imaginary parts of $b_k$ are
\begin{eqnarray}
    \mathrm{Re}(b_k) &=& \ell \left(\frac{\partial \tilde{\omega}}{\partial \ell} \right)^{-1}_n 
    \sum_{s+k = \mathrm{odd}} (-1)^{\frac{s-k+1}{2}}
    \frac{(s-k)!!(s+k)!!}{(s+k)!} \times 
    P_s^k \left(\frac{m}{\ell} \right) 
    \langle \Omega_s \rangle_{n\ell}, 
    \quad k = 1,2,... \label{eqn: b_k_real} \\
    \mathrm{Im}(b_k) &=& k\ell 
    \left(\frac{\partial \tilde{\omega}}{\partial \ell} \right)^{-1}_n
    \sum_{s+k = \mathrm{even}} (-1)^{\frac{s-k+2}{2}}
    \left(\frac{2s+1}{4\pi}\right)^{1/2} \frac{(s-k-1)!!(s+k-1)!!}{(s+k)!} \times P_s^k \left(\frac{m}{\ell} \right) \langle \frac{v_s}{r} \rangle_{n\ell}, \quad k = 1,2,... \label{eqn: b_k_imag}.
\end{eqnarray}
We consider only odd-$s$ dependencies of $\Omega$. The even-$s$ correspond
to North-South (NS) asymmetry in differential rotation and are estimated
to be weak at the surface \citep[NS asymmetry coefficients are estimated to be an order of magnitude smaller than their symmetric counterparts;][]{mdzinarishvili2020}.
The contribution of even-$s$ components to the real part of $b_k$
can thus be ignored. For the asymptotic limit of high-degrees,
\begin{equation} \label{eqn: b_k_odd_s}
    \mathrm{Re}(b_k) = \ell \left(\frac{\partial \tilde{\omega}}{\partial \ell} \right)_n^{-1} \sum_{s+k = odd} (-1)^{\frac{k-2}{2}} \frac{s!(s-k)!!(s+k)!!}{(s+k)!s!!s!!} \times a^{n\ell}_s \, P_s^k \left(\frac{m}{\ell} \right), \quad k = 2,4,... ,
\end{equation}
\begin{equation}
    a^{n\ell}_s \approx (-1)^{\frac{s-1}{2}} \frac{s!! s!!}{s!} \langle \Omega_s\rangle_{n\ell}, \quad s = 1,3,...
\end{equation}

Figure~\ref{fig:efn_pert} illustrates the distortion of eigenfunctions due to an equatorially symmetric 
differential rotation (using frequency splitting estimates of $a_3$ and $a_5$ coefficients). 
It can be seen that differences between distorted
eigenfunctions $\vec{\xi}_{nlm}$ and their undistorted counterparts 
$\tilde{\vec{\xi}}_{nlm}$ are at around the $50\%$ level for some azimuthal
orders. The correction coefficients, given by $c^{\ell+p, m}_{\ell, m}$, are
shown in the bottom panel of Figure~\ref{fig:efn_pert}. Since the largest contribution to 
$\xiv_{\ell}$ comes from $\tilde{\xiv}_{\ell}$, $c^{\ell, m}_{\ell, m} (\gtrsim 0.8)$ are not plotted to
highlight the corrections from neighbouring modes with $p \neq 0$. Visual inspection shows that
$c^{\ell+p, m}_{\ell, m}$ have non-zero elements at $p = \pm 2, \pm 4$, as expected
from selection rules due to the rotation field $\Omega_s(r)$ for $s=3,5$. High $\ell$
eigenfunctions are predominantly large close to the surface. Consequently, we see
that their distortions are much larger at shallower than deeper depths. We choose to plot
three cases --- low, intermediate, and high $m$. For the extreme cases of $m = 0$ and $m = \ell$, 
$c^{\ell+p, m}_{\ell, m} \sim 0$, since for odd $s$ and even $k$, $P_s^k(\mu)$ vanishes at $\mu = 0,1$. Thus these eigenfunctions remain undistorted under an equatorially symmetric differential rotation. 

For sake of completeness, it may be mentioned that the finite $c^{\ell+p, m}_{\ell, m}$ for $p \neq 0$
seemingly disqualifies the frequency-splitting measurements, which assume isolated multiplets --- meaning 
$c^{\ell+p, m}_{\ell, m} = \delta_{p,0}$. However, it does not necessarily imply that the isolated multiplet 
approximation is poor at these angular degrees. If the eigenfunction error $\delta \xiv_k$ incurred on neglecting cross-coupling is of
order $\mathcal{O}(\epsilon)$ then it can be shown \citep[see Chapter 8 of][]{freidberg_2014, cutler} that the error in estimating eigenfrequency $\delta \omega_k$ is at most of order $\mathcal{O}(\epsilon^2)$, where $\epsilon$ is small. To illustrate this further, if the error in estimating eigenfunction distortion on neglecting cross coupling $(p \neq 0)$ is written as $\epsilon \, \xiv_{\ell + p}$, then from inspecting Eqn.~(\ref{eqn:voront-pert}), we see that $\epsilon \sim |c_{\ell}^{\ell+p}|$. Upon investigating the $(n,\ell) = (2,150)$ case presented in Figure~\ref{fig:efn_pert} for $p \neq 0$, we find $c_{\ell}^{\ell+p} \lesssim \mathcal{O}(10^{-1})$. The equivalent error incurred in eigenfrequency estimation may be computed according to the discussion in Section~\ref{sec:QDPT_vs_DPT}. This yields $\delta \omega / \omega \lesssim \mathcal{O}(10^{-2})$ in the range $150 \leq \ell \leq 250$ thereby confirming the above argument for $\epsilon \sim 10^{-1}$.
Given the leakage matrices and Lorentzians, the forward
problem of modeling $\langle \varphi^{\ell' m'} \varphi^{\ell m *} \rangle$
requires constructing eigenfunction corrections $c^{\ell+p}_{\ell}$ 
using the $a$-coefficients in Eqn.~(\ref{eqn: b_k_odd_s}) and the poloidal
flow in Eqn.~(\ref{eqn: b_k_imag}). Thus, for axisymmetric flows, the cross
spectra for moderately large $\ell_1$ and $\ell_2$ from 
Eqn.~(\ref{eqn: mode_coupling}) may be written more explicitly as
\begin{equation}
    \langle \varphi^{\ell_1, m_1} \, \varphi^{\ell_2, m_1 *} \rangle =
    \sum_{p,p',\ell,m} \, \tilde{L}_{\ell+p,m}^{\ell_1,m_1} \,
    \tilde{L}_{\ell+p',m}^{\ell_2, m_2} \, c_{\ell,m}^{\ell+p,m} \,
    c_{\ell,m}^{\ell+p',m*} \, \langle |\Lambda^{\ell,m}(\omega)|^2 \rangle.
\end{equation}
 The leakage matrices $\tilde{L}_{\ell+p,m}^{\ell_1,m_1}$ impose bounds on the farthest modes that leak into mode amplitude $\varphi^{\ell m}$. This is because $\tilde{L}_{\ell+p,m}^{\ell_1,m_1}$ is non-zero only when $\ell+p \in [\ell_1 - \delta \ell, \ell_1 + \delta \ell]$ and 
 $m \in [m_1 - \delta m, m_1 + \delta m]$, where $\delta \ell$ and
 $\delta m$ are the farthest spectral offsets.
 Thus, for a given $\ell$, we must
 determine the correction coefficients $c_{\ell,m}^{\ell+p,m}$ such 
 that $p \in [\ell_1 - \delta \ell -\ell,\ell_1 + \delta \ell-\ell]$. 
 Similar bounds on $p'$ in $c_{\ell,m}^{\ell+p',m}$ are imposed by the second leakage matrix $\tilde{L}_{\ell+p',m}^{\ell_2,m_2}$,
 namely,
\begin{equation}
    \langle \varphi^{\ell_1,m_1} \, \varphi^{\ell_1 + \Delta \ell, m_1 *} \rangle = 
    \sum_{p,p',\ell,m} \, \tilde{L}_{\ell+p,m}^{\ell_1,m_1} \, 
    \tilde{L}_{\ell+p',m}^{\ell_1 + \Delta \ell, m_1} \, 
    c_{\ell,m}^{\ell+p,m} \, c_{\ell,m}^{\ell+p',m*} \, \langle |\Lambda^{\ell,m}(\omega)|^2 \rangle
    .
    \label{eqn:mode-coupling}
\end{equation}
Being 
significantly weaker than  differential rotation, we neglect the contribution 
of meridional circulation \citep{imada18,gizon2020_sci} to the eigenfunction corrections.

 \section{Data Analysis}
 \label{sec:data_analysis}
 
 \begin{figure}
     \centering
     \includegraphics[width=1.0\textwidth]{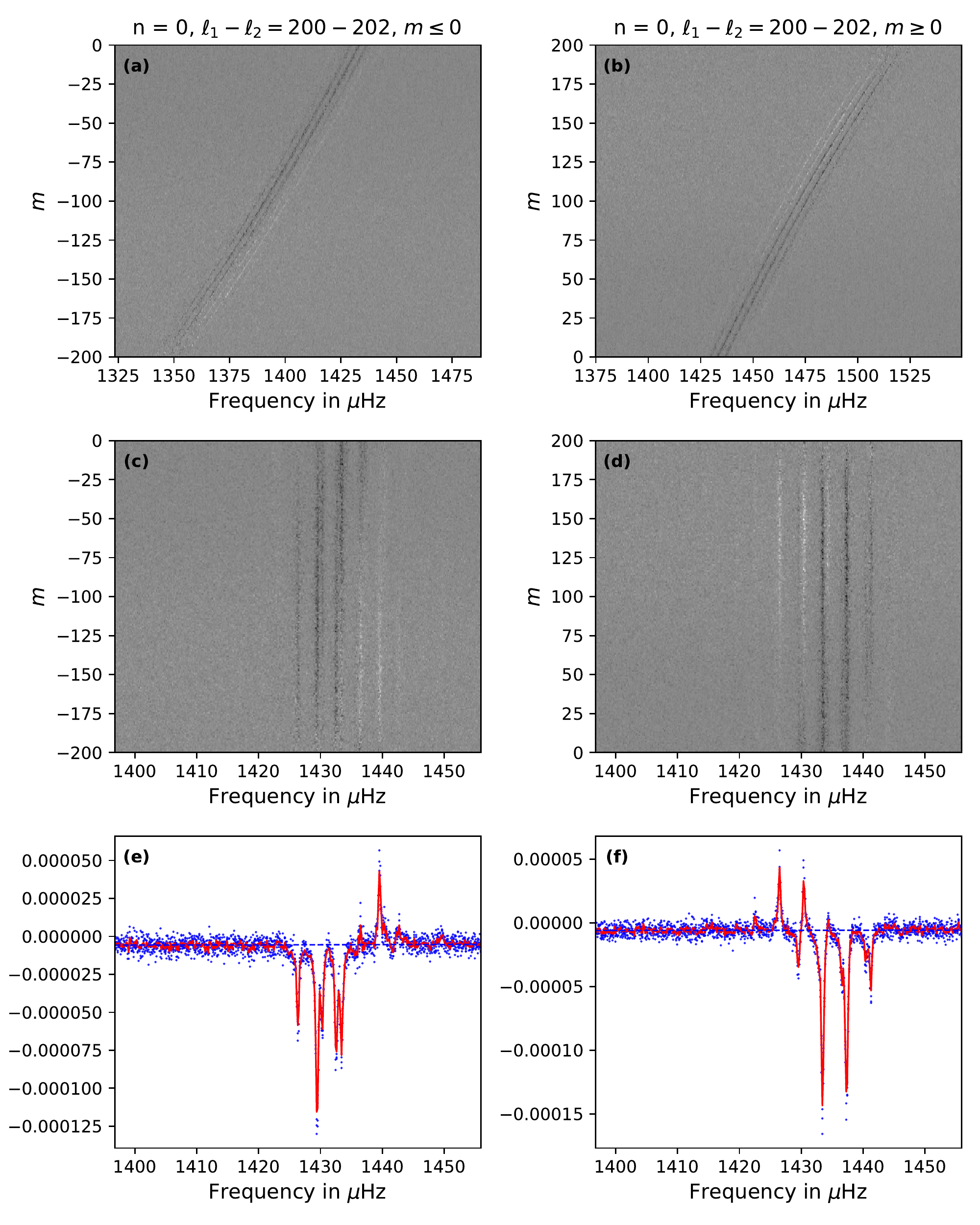}
     \caption{Cross-spectral signal for $\ell = 200$, $\Delta \ell = 2$ 
     and $n=0$. 
     Panel (a, b): 
     Observed cross-spectrum corresponding to $m^+$ and $m^-$. 
     Panel (c, d): Derotated cross spectrum corresponding to $m^+$ and $m^-$.
     Panel (e, f): $D^{\ell, \Delta \ell, \pm}_n$. The baseline is 
     indicated by the dashed blue line. The blue dots represent 
     observations from the five 72-day time series and the red curve
     corresponds to the expectation value of the cross-spectrum.}
     \label{fig:cs-200-202}
 \end{figure}
 
 We use the full-disk 72-day gap-filled spherical-harmonic time 
 series $\varphi^{\ell m}(t)$, which are recorded at a cadence 
 of 45~seconds by HMI \citep{larson-schou-2015}.
 The data are available for harmonic degrees 
 in the range $\ell \leq 300$.
 The time series is transformed to the frequency domain to obtain 
 $\varphi^{\ell m}(\omega)$. 
 The negative-frequency components are associated
 with the negative $m$ components using the symmetry relation 
 (Appendix~\ref{apdx:sph-symm})
 \begin{equation}
     \varphi^{\ell, -|m|}(\omega) = (-1)^{|m|}\varphi^{\ell, |m|*}(-\omega).
 \end{equation}
 The ensemble average of the cross spectrum  is computed by averaging 
 five continuous 72-day time series, which corresponds to 360~days of 
 helioseismic data. The eigenfrequencies of the unperturbed model 
 $\tilde{\omega}_{n\ell m}$ are degenerate in~$m$, i.e., 
 $\tilde{\omega}_{n\ell m} = \tilde{\omega}_{n\ell 0}$.
 Rotation breaks spherical symmetry and lifts the degeneracy in~$m$.
 As in W13, we show the cross-spectrum for $n=0$ and 
 $\ell=200$, $\Delta \ell=2$ in Figure~\ref{fig:cs-200-202}.
 The effect of rotation is visible through the inclination of the ridges in the $m-\nu$
 spectrum, as seen in Panels~(a, b) of the figure.
 The multiple vertical ridges are due to leakage of power.

 The cross spectra are derotated and stacked about the central frequency, corresponding to~$m=0$, which is shown in Panels~(c, d) of 
 Figure~\ref{fig:cs-200-202}. In order to improve the signal-to-noise
 ratio, the stacked cross spectrum is summed over azimuthal order~$m$.
 This quantity is used to determine the extent of coupling,
 denoted by $D^{\ell,\Delta \ell, \pm}_n$. 
 The~$-$~($+$) signs indicates summation over negative (positive)~$m$. 
 For notational convenience, we define~$m^+$ when 
 referring to $m \geq 0$ and $m^-$ to denote $m \leq 0$.
 The operation of stacking (derotating) the original spectra is denoted by $\stackm$. 
 Since differential rotation affects only the real part of the
 cross-spectrum (see Eqn.~\ref{eqn:clp}), $D^{\ell, \Delta \ell, \pm}_n$ refers to 
 the real part of the cross-spectrum.
 \begin{equation}
     D_n^{\ell, \Delta \ell, \pm}(\omega) = \left \langle \sum_{m^\pm} \stackm \left( 
     \text{Re}\left[\varphi^{\ell m}(\omega) \varphi^{\ell+\Delta \ell, m*}(\omega) 
     \right] \right) \right\rangle .
     \label{eqn:data-measurement}
 \end{equation}
  The cross-spectral model is a combination of Lorentzians and is
  based on Eqn.~(\ref{eqn:mode-coupling}).
  The HMI-pipeline analysis provides us with mode amplitudes and
 linewidths for multiplets $(n, \ell)$. The~$m$ dependence of 
 frequency, $\omega_{n\ell m} - \omega_{nl0}$, is encoded in 36
 frequency-splitting coefficients $(a^{nl}_1, a^{nl}_2, ..., a^{nl}_{36})$.
 These values are used to construct the Lorentzians for the model,
which is denoted by $M^{\ell, \Delta \ell, \pm}$ and expressed as
\begin{equation}
    M^{\ell, \Delta \ell, \pm}_n(\omega) =  \sum_{m\pm} \stackm \left(
    \sum_{p,p',\ell,m'} \, \tilde{L}_{\ell+p,m'}^{\ell_1,m} \, 
    \tilde{L}_{\ell+p',m'}^{\ell_1 + \Delta \ell, m} \, 
    c_{\ell,m'}^{\ell+p,m'} \, c_{\ell,m'}^{\ell+p',m'*} \, \langle |\Lambda^{\ell,m'}_n(\omega)|^2 \rangle
    \right)
    \label{eqn:mode-coupling-model}.
\end{equation}
 As seen in Panels~(e, f) of Figure~\ref{fig:cs-200-202}, the cross spectra
 sit on a non-zero baseline. This is a non-seismic background and hence is explicitly fitted for before
 further analysis of the data. The complete model of the cross spectrum
 involves leakage from the power spectrum, eigenfunction coupling, as 
 well as the non-seismic background, i.e., the data $D^{\ell, \Delta\ell, \pm}_n(\omega)$
 is modelled as 
\( M^{\ell, \Delta \ell, \pm}_n(\omega) + b^{\ell, \Delta \ell, \pm}_n (\omega) \).
The baseline $b_n^{\ell, \Delta \ell, \pm}(\omega)$ is computed by considering 
50 frequency bins on either side, far from resonance, and 
fitting a straight line through them, in a least-squares sense. 
The model $M^{\ell, \Delta \ell, \pm}_n(\omega)$ 
depends on the $a^{nl}_3$ and $a^{nl}_5$ splitting coefficients 
via the eigenfunction-correction coefficients $c^{\ell + p}_{\ell}$.
A Bayesian-analysis approach is used to estimate the values 
 $(a^{nl}_3, a^{nl}_5)$, using MCMC, described in Section~\ref{sec:MCMC}.
The misfit function that quantifies the goodness of a chosen model
is given by
\begin{equation}
    \Xi_n = \sum_{l, \omega, \pm} \left( \frac{ D^{\ell, \Delta \ell, \pm}_n(\omega)
    - (M^{\ell, \Delta \ell, \pm}_n (\omega) + b^{\ell, \Delta \ell, \pm}_n(\omega))}
    {\sigma^{\ell, \Delta \ell, \pm}_n(\omega)} \right)^2
    \label{eqn:misfit},
\end{equation}
where $[\sigma^{\ell, \Delta \ell, \pm}_n(\omega)]^2$ denotes the variance of the 
data $D^{\ell, \Delta \ell, \pm}_n(\omega)$ and is given by
\begin{equation}
   [\sigma^{\ell, \Delta \ell, \pm}_n(\omega)]^2 = \left\langle \left(
   \sum_{m\pm} \stackm \left[\phi^{\ell, m}(\omega)\, \phi^{\ell+\Delta \ell, m*}(\omega) \right] - 
   D^{\ell, \Delta \ell, m\pm} \right)^2 \right\rangle .
   \label{eqn:variance}
\end{equation}

\subsection{Bayesian Inference: MCMC} \label{sec:MCMC}
Bayesian inference is a statistical method to determine the probability
distribution functions (PDF) of the inferred model parameters. 
For data $D$ and model parameters $a$, the posterior PDF $p(a|D)$, which 
is the conditional probability of the model given data, 
may be constructed using the likelihood function $p(D|a)$ and 
a given prior PDF of the model 
parameters $p(a)$. 
The prior encapsulates information about what is 
already known about the model parameters $a$.
\begin{equation}
    p(a|D) \propto p(D|a) p(a).
\end{equation} 
The constant of proportionality is the normalization factor 
for the posterior probability distribution, 
which may be difficult to compute. The sampling of these PDFs is performed
using MCMC, which involves performing a biased random walk
in parameter space. Starting from an initial guess of parameters, 
a random change is performed. The move is accepted or rejected
based on the ratio of the posterior probability at the two locations.
Hence, the normalization factor is superfluous to the MCMC method.

Bayesian MCMC analysis has been used quite extensively in astrophysical 
problems 
\citep[][and references therein]{saha-williams-1994,christensen-meyer-1998, sharma2017mcmc} and terrestrial seismology 
\citep[][and references therein]{sambridge2002}. 
However, the use of MCMC in global helioseismology has been limited as 
compared to terrestrial seismology \citep{jackiewicz2020}.

The aim of the current calculation is the estimation of 
$(a_3^{n\ell}, a_5^{n\ell})$ that best reproduce the observed 
cross-spectra from the model, given by 
Eqn.~(\ref{eqn:mode-coupling-model}), where it is seen that
the coupling coefficients $c^{\ell + p}_{\ell}$ depend on
$(a_3^{n\ell}, a_5^{n\ell})$. However, because of leakage, neighbouring
$\ell$ corresponding to the spectrum in question also contribute
to the cross-spectrum. Hence, the spectrum of $(\ell, \Delta \ell)$
depends on $(a_3^{n\ell'}, a_5^{n\ell'})$ for
$\ell' \in [\ell-\delta\ell, \ell+\Delta\ell+\delta\ell]$. 
Since we only consider mode leakage  at the same
radial order $n$, we are forced to simultaneously estimate all 
the $(a_3^{n\ell}, a_5^{n\ell})$ for a given $n$. For instance,
at $n=0$, we have 52 modes with $\ell < 250$, and 94 spectra 
corresponding to $\Delta \ell = 2, 4$, for both $m^+$ and $m^-$
branches. In this case, there are 52 $(a_3^{0\ell}, a_5^{0\ell})$ pairs
that need to be estimated and 188 spectra which need to be modeled.
Performing inversions on a high dimensional, jagged landscape is 
a challenge as the fine tuning of regularization is tedious. 
However, since we have a model which encodes the dependence of the
$a$ coefficients on the cross-spectrum, we could ``brute-force'' the
estimation of parameters. The utility of MCMC is that it
enables us to sample the entire parameter space. 
Since the inference of the posterior PDF depends strongly on the prior, it is instructive to use
an uninformed or flat prior. 

For the MCMC simulations, we use the Python package \texttt{emcee} by \cite{emcee}.
The package is based on the affine invariant ensemble sampler 
by \cite{goodman_weare_2010}. Multiple random walkers are used 
to sample high-dimensional parameter spaces efficiently. We use a 
flat prior for all $a_3^{n\ell}$ and $a_5^{n\ell}$ given by

\begin{align}
    p(a_3) = \frac{1}{20} \qquad 15 \le  a_3 \le 35 
    \qquad \text{and} \qquad 
    p(a_5) = \frac{1}{16} \qquad -16 \le  a_5 \le 0,
\end{align} 
and zero everywhere else
for all $(\ell, n)$. This is motivated by the results of frequency
splittings. For modes near the surface, i.e., for low values of 
$\nu_{n\ell}/\ell$, $a_3$ has been measured to be nearly $22$~nHz
and $a_5$ is $-4$~nHz. The likelihood function is defined as
\begin{equation}
    p(D|a) = \exp(-\Xi_n), \label{eqn:likelihood}
\end{equation}
where~$\Xi_n$ is the misfit given by Eqn.~(\ref{eqn:misfit}). Flat priors
enable us to sample the likelihood function in the given region in 
parameter space. We perform MCMC inversions for $n=0, 1,..8$ and find that the likelihood function is unimodal in all model parameters.
For the sake of illustration, a smaller computation is presented in 
Appendix~\ref{sec:MCMC_demo}. 

  \begin{table}[h!]
  \begin{varwidth}[b]{0.35\linewidth}
    \centering
    \begin{tabular}{| c | c |}
     \hline
     Radial order $n$ & Range of $\ell$ for $(a_3, a_5)$\\
     \hline\hline
     0  &  192--241, 241--281, 271--289 \\
     1  &  80--120, 110--150, 140--183 \\
     2  &  60--100, 90--130, 120--161 \\
     3  &  43--73, 73--113, 103--145 \\
     4  &  40--80, 70--110, 100--140 \\
     5  &  46--86, 76--116, 106--146 \\
     6  &  58--98, 88--128, 118--138 \\
     7  &  64--104, 94--114 \\
     8  &  73--103 \\
     \hline
    \end{tabular}
    \vspace*{7mm}
    \caption{List of modes $(n,\ell)$ used in MCMC. These
    are marked as \textit{black} dot in Figure~\ref{fig:mode-selection}.}
    \label{tab:modelist}
  \end{varwidth}%
  \hfill
  \begin{minipage}[b]{0.65\linewidth}
    \centering
    \includegraphics[width=0.95\textwidth]
    {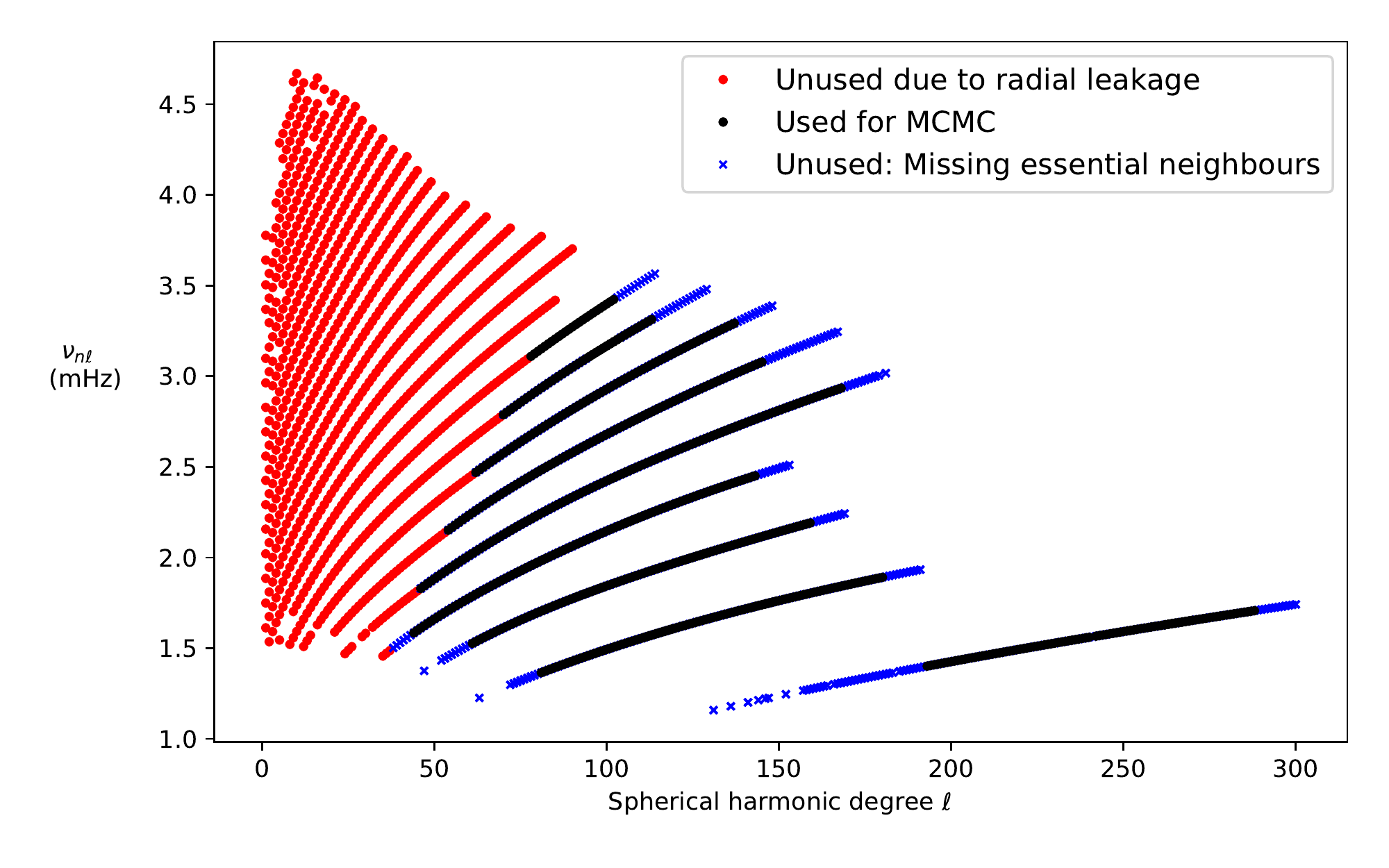}
    \captionof{figure}{Classification of modes.}
    \label{fig:mode-selection}
  \end{minipage}
\end{table}

 \section{Results and Discussion} \label{sec:results}
 
 The MCMC analysis is performed for each radial order separately. 
 The current model only considers leakage between modes of the
 same radial order and hence the ideal way of estimating the parameters would
 be to estimate all $(a_3, a_5)$ at a given radial order by modelling
 all the cross-spectra at the same radial order. However, this makes the
 problem computationally very demanding as the MCMC method used requires 
 at least $2k+1$ random walkers for $k$ different parameters to be fit. To
 work around this, we break the entire set of parameters into chunks of 
 40 pairs, while ensuring an overlap of 10 pairs between the chunks. In 
 Table~\ref{tab:modelist}, we list the set of $\ell$'s for which MCMC
 sampling is performed and parameters are estimated. 
 
 Figure~\ref{fig:mode-selection} marks the multiplets $(n,\ell)$ available from the HMI pipeline. 
 The multiplets whose modes are used for this study are labelled as black dots. 
 The red dots, which are located at
 lower $\ell$, correspond to those modes which have contributions
 from neighbouring radial orders within the temporal-frequency window. This
 gets worse for $\ell<20$, where contributions from neighbouring radial orders
 may be seen even near central peaks. Modelling these spectra would
 require including coupling across radial orders, which is not the case in the present analysis. Thus we only use modes corresponding to $\nu_{n\ell}/\ell < 45$. Figure~\ref{fig:mode-selection}
 also marks unused HMI-resolved modes as blue dots on 
 either side of the black dots (used modes). This is because we consider
 only modes that may be fully modelled with parameters available from the HMI pipeline. Modelling a the degree $\ell$ requires
 mode parameters corresponding to modes from $(\ell-\delta\ell)$ to 
 $(\ell+\delta\ell)$. The existence of unresolved modes (with no mode-parameter 
 information from the HMI pipeline) in this region
 means that modelling is incomplete, i.e., there would be peaks in the
 observed spectrum that are missed by the model. Hence, such modes are
 not considered for the present work. For any given radial order, the first
 $\delta\ell$ and the last $\delta\ell$ modes cannot be modelled and thus
 we see blue points on either side of the set of black dots in 
 Figure~\ref{fig:mode-selection}. 
 
 The results of the MCMC analysis at all the radial orders are combined
 and presented in Figure~\ref{fig:error-bars}. We note that that
 the confidence intervals become larger for higher $\nu/\ell$. The reasons
 for this are discussed in Section~\ref{sec:a-coeff-sens}.  Estimates 
 of $a$-coefficients are largely in agreement with the splitting
 coefficients --- although the most probable values of the coupling-derived parameters are different from their splitting counterparts, they predominantly lie
 within the 1-$\sigma$ confidence interval. The confidence intervals of
 $a_3$ and $a_5$ are nearly the same size.
We obtain better results, in terms of the spread in the inferred
$a_3$-coefficients, than W13. 
This may be attributed to the consideration of data variance as well
as simultaneous fitting for model parameters using a Bayesian approach. 
For instance, the spread of $a_3$ in the range $0 < \nu/\ell \lesssim 40$ 
is seen to be in the range 7.5--30~nHz in W13,
whereas our estimates are in the range 15--26~nHz. The present method
allows us to quantify the 1-$\sigma$ confidence interval around the most probable
values for estimated $a$-coefficients, whereas 
W13 have shown only inversion values of $a$-coefficients without their
respective uncertainties. However, we also note that the estimates of $a_5$
from Bayesian analysis are comparable to the least-squares inversions of W13.

 \subsection{Reconstructed power and cross spectra}
 \label{sec:reconst-spectra}
 
 \begin{figure}[h!]
    \centering
    \includegraphics[width=\textwidth]{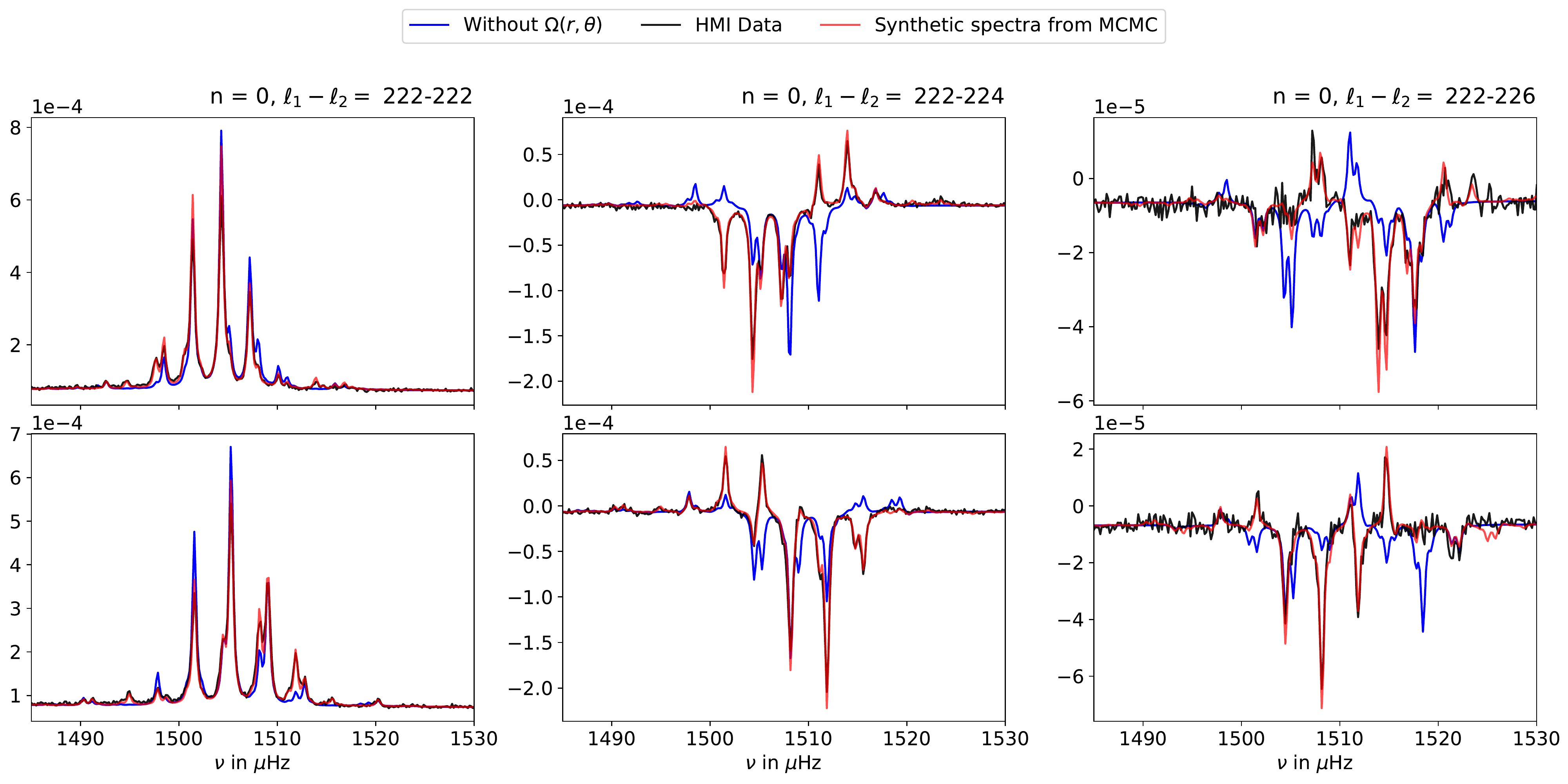}
    \caption{Cross spectrum for $\ell=222$ and $\Delta\ell = 0, 2, 4$. 
    The upper panels correspond to $m^+$ and lower panels  
    to $m^-$.
    The black curve shows observed data. The blue curve is the model
    before considering eigenfunction coupling and the red curve corresponds
    to model constructed using parameters estimated from MCMC.}
    \label{fig:spectra_00_222}
\end{figure}

\begin{figure}[h!]
    \centering
    \includegraphics[width=\textwidth]{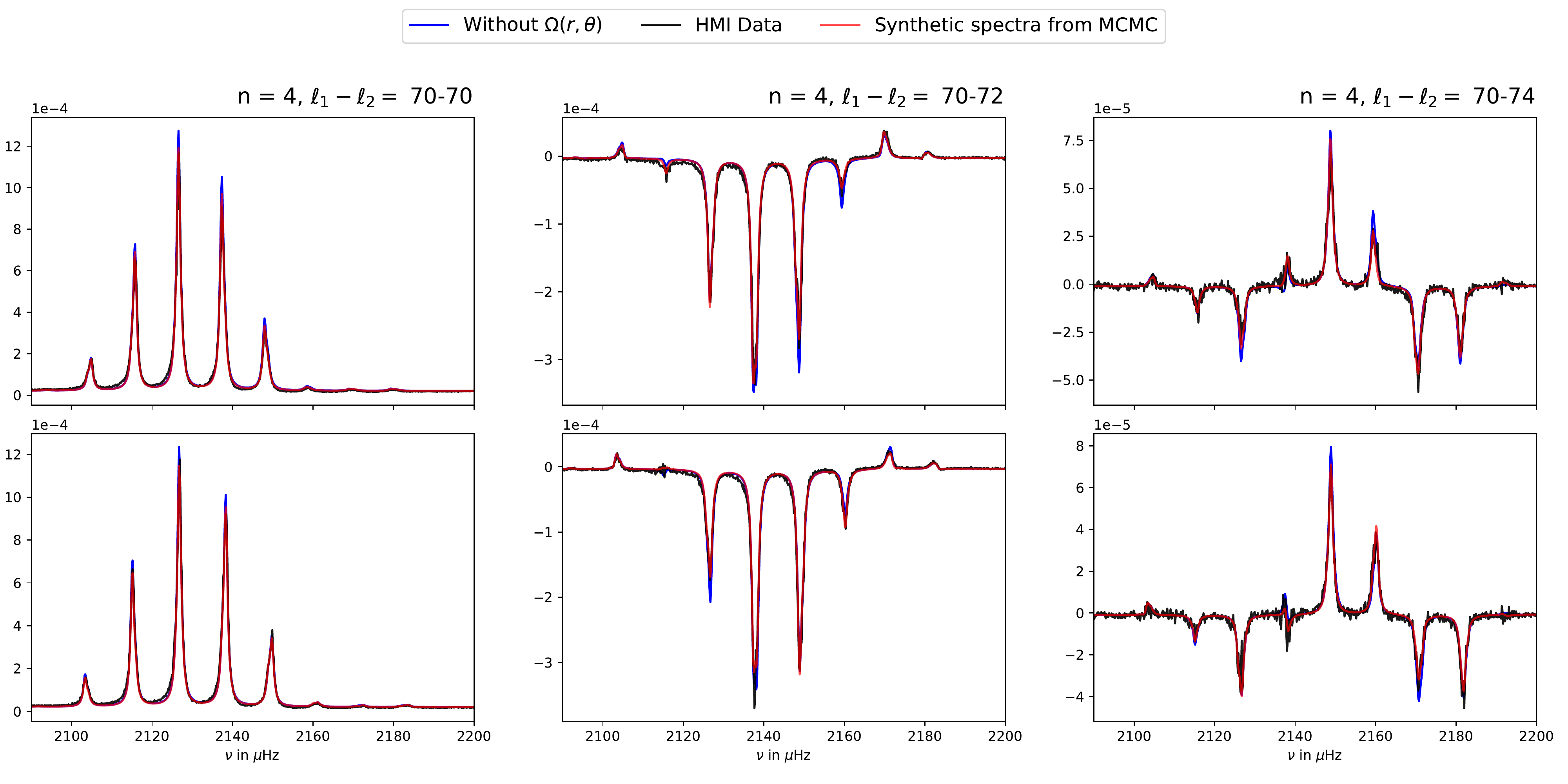}
    \caption{Cross spectrum for $\ell=70$ and $\Delta\ell = 0, 2, 4$. 
    The upper panels correspond to $m^+$ and lower panels  
    to $m^-$.
    The black curve shows observed data. The blue curve is the model
    before considering eigenfunction coupling and the red curve corresponds
    to the model constructed using parameters estimated from MCMC.}
    \label{fig:spectra_04_70}
\end{figure}
 
 The $a$-coefficients obtained from the MCMC analysis are used to reconstruct
 cross-spectra, e.g., Figure~\ref{fig:spectra_00_222} shows the cross spectrum
 for $(n=0, \ell=222)$. It may be seen that, before considering
 eigenfunction corrections (in the absence of differential rotation), 
 the spectrum shown in blue is considerably
 different --- in both magnitude
 and sign --- from the observed data. After including eigenfunction corrections,
 which have been estimated from MCMC, we see that the model is in close
 agreement with the data. In the intermediate-$\ell$ range, we show 
 cross-spectra for $(n=4, \ell=70)$ in Figure~\ref{fig:spectra_04_70}. The corrections due to eigenfunction distortion are markedly
 less significant when compared to $(\ell=222, n=0)$, demonstrating loss
 of sensitivity of the model to the coupling coefficients.

\begin{figure}
    \centering
    \includegraphics[width=1.0\textwidth]{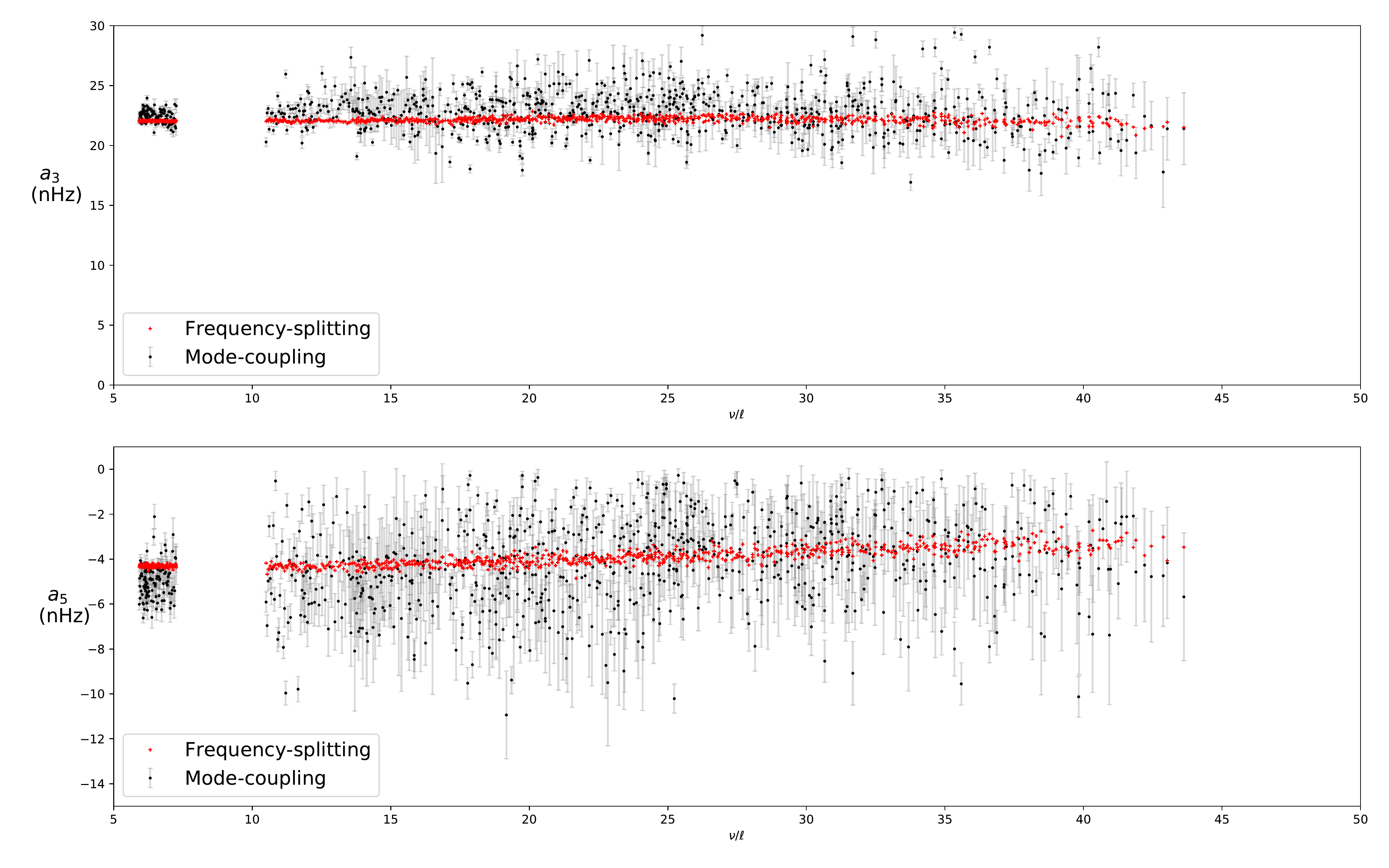}
    \caption{Inferred $a_3$ and $a_5$ coefficients from MCMC
    are shown as black dots with 1-$\sigma$ confidence intervals.
    The values from frequency splitting
    are shown in red.}
    \label{fig:error-bars}
\end{figure}

\subsection{Sensitivity of $a$-coefficients to differential rotation}
\label{sec:a-coeff-sens}

Mode coupling has diminished sensitivity in estimating $a$-coefficients for low-$\ell$ modes. The coupling coefficients
$c_\ell^{\ell+p}$ depend on the real and imaginary parts of $b_k$. 
Differential rotation contributes to only the real part of $b_k$
(Eqn.~[\ref{eqn: b_k_real}]) and the dependence
on $\ell$ appears through the factor
$\ell (\partial \omega_{nl}/\partial \ell)^{-1}$. 
The plot of eigenfrequencies $\omega_{n\ell}$ against $\ell$ is 
known to flatten for higher $\ell$. Hence,
$\partial \omega_{n\ell}/\partial \ell$ is large for small $\ell$ and
small for large $\ell$ \citep[see Figure~1 in][]{rhodes1997}. This results in $b_k$ being small for low $\ell$ and its magnitude increases with $\ell$, causing this decreased
sensitivity to low $\ell$. 
The lower sensitivity implies that the misfit function $S$ is flatter 
at lower $\ell$. To demonstrate this, we compute $S$ over $\ell=80$--245 for a range of values of $a-$coefficients and determine 
how wide or flat $S$ is in the neighbourhood of the optimal solution.

\begin{figure}
    \centering
    \includegraphics[width=\textwidth]{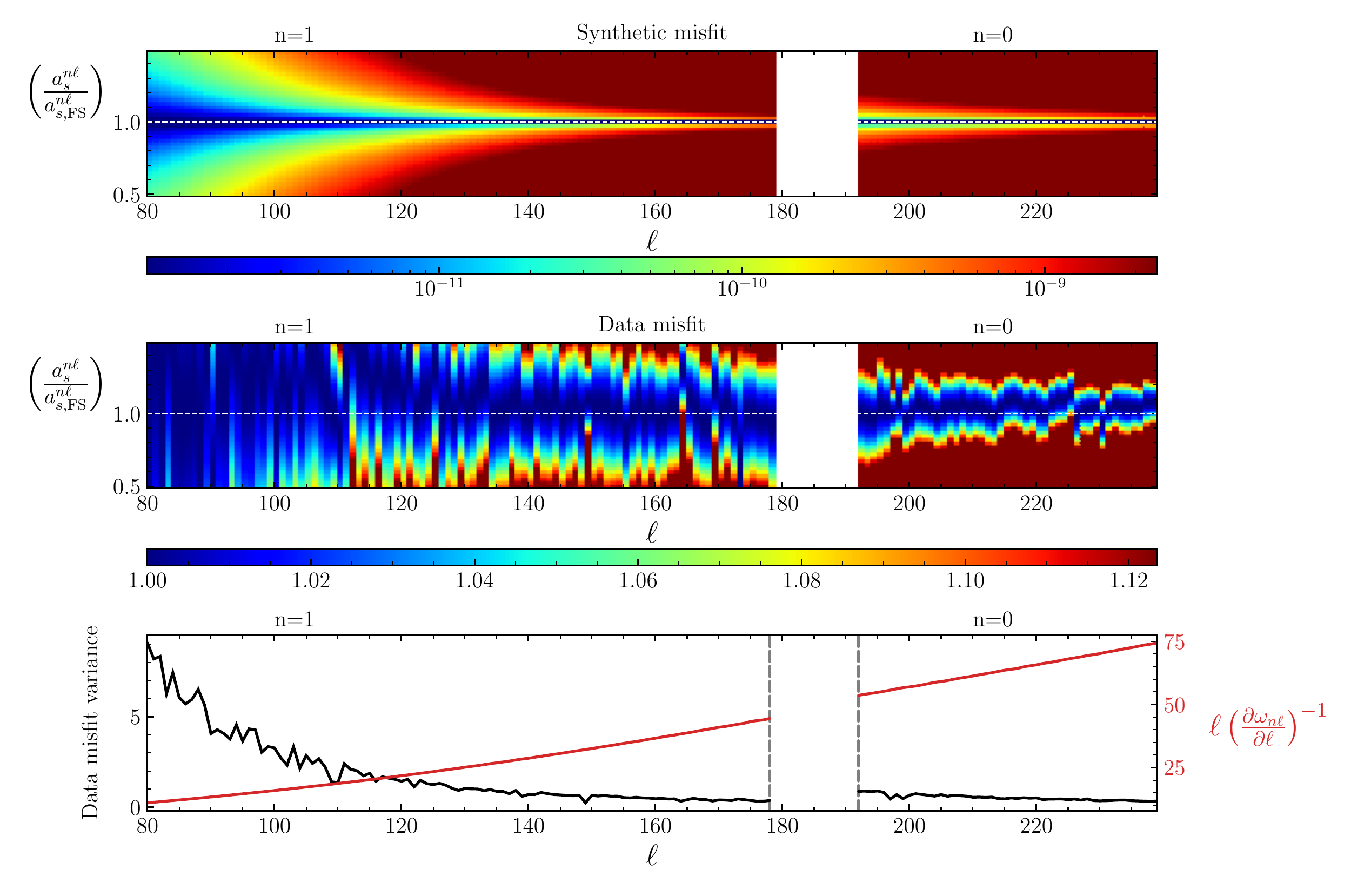}
    \caption{Sensitivity of spectral fitting to $a$-coefficients as a
    function of angular degree $\ell$. \textit{Top} panel shows the 
    variation of misfit between synthetic data calculated using 
    frequency-splitting $a$-coefficients $a_{s,\mathrm{FS}}^{n\ell}$ 
    and synthetic spectra computed from a scaled set 
    of $a$ coefficients $a_{s}^{n\ell}$. A well-defined minimum along 
    $a_s^{n\ell}/a_{s,\mathrm{FS}}^{n\ell} = 1.0$, which broadens 
    towards smaller $\ell$, shows a drop in sensitivity of the spectra 
    to variations in $a$ coefficients, as predicted by theory. \textit{Middle} panel shows the sensitivity of
    $a$ coefficients, but now computed using the misfit between HMI and synthetic spectra computed from a scaled 
    set of $a$ coefficients $a_{s}^{n\ell}$. While it has the same
    qualitative drop in $a$-coefficient sensitivity for decreasing
    $\ell$, the ridge of the minimum (darkest patch) is seen to 
    deviate from $a_{s,\mathrm{FS}}^{n\ell}$. 
    \textit{Bottom} panel shows in \textit{black} the effective 
    variance of misfit for each $\ell$. 
    The narrowing confinement of the data misfit towards higher 
    $\ell$ is seen as a decreasing effective variance with 
    increasing $\ell$. The \textit{red} line shows the increase 
    in the factor $\ell/(\partial \omega_{n\ell}/\partial \ell)$ 
    that enhances sensitivity at higher $\ell$, as predicted by
    Eqn.~(\ref{eqn: b_k_odd_s}). The areas corresponding to 
    radial orders $n=0, 1$ are indicated on top of each plot.
    }
    \label{fig:acoef_sens}
\end{figure}

Figure~\ref{fig:acoef_sens} shows that the misfit is 
wide for $\ell=80$ and it becomes sharper with increasing $\ell$. As the
highest-resolved mode for $n=1$ corresponds to $\ell=179$, we consider
the radial order $n=0$ in order to study this in an extended region of 
$\ell$. The first two panels show the colour map of the misfit function.
Near the optimal value $a^{n\ell}_s/a^{n\ell}_{FS} = 1$, the synthetic misfit falls to~$0$.
This is possible as the synthetic data is noise free and it can be completely modeled. The misfit increases on either side of the optimum
value. The second panel shows the scaled misfit for HMI data, which
is close to $1$ at the optimum, increasing on either side of the optimal
value. We see the dark patch become wider at lower $\ell$, indicating the
flatness of the misfit function for low $\ell$. The likelihood function,
which is defined to be $\exp(-S)$, is approximated as a Gaussian
in the vicinity of the optimum. The width of this Gaussian
is treated as a measure of the width of the misfit function $S$, with wider
misfit implying lower sensitivity to $a-$coefficients.
This is shown in the third panel of Figure~\ref{fig:acoef_sens}, 
where we see a decreasing trend in misfit width, indicating that the
sensitivity of mode coupling increases with $\ell$.

\subsection{Scaling factor for synthetic spectra}

\begin{figure}
    \centering
    \includegraphics[width=\textwidth]{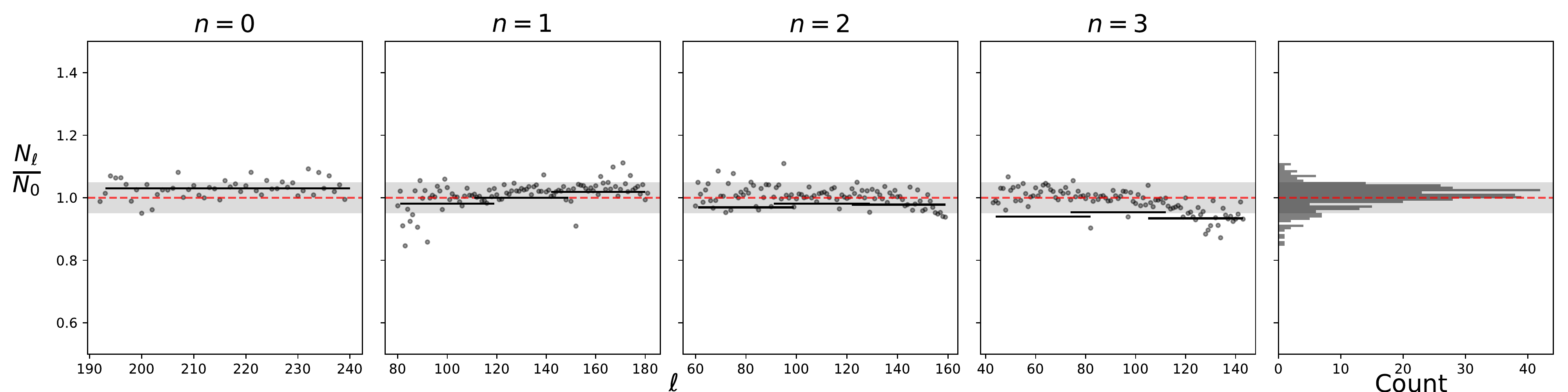}
    \caption{The red line corresponds to $N_0$. The gray region corresponds
    to 5\% error from $N_0$. The gray points correspond to $N_\ell$ and the
    solid black lines are from each MCMC simulation. The right-most panel
    shows the histogram of all the gray points, taken from all radial orders.}
    \label{fig:norm_plot}
\end{figure}

The model constructed using mode parameters obtained from the HMI 
pipeline needs to be scaled to match the observations. This scaling 
factor has to be empirically determined. Since there is no well-accepted
convention to estimate this factor, it is worthwhile to explore different
methods of its estimating. We employ three different methods to
infer the scale factor and show that the results are nearly identical.
\begin{compactitem}
    \item Consider all the power spectra for a given radial order
    and perform a least-squares fitting for the scale factor $N_0$.
    \item Fit for the scale factor $N_\ell$ as a function of the spherical
    harmonic degree $\ell$ by considering all power spectra 
    at a given radial order.
    \item Include the scale factor as an independent parameter to be 
    estimated in the MCMC analysis.
\end{compactitem}
 Figure~\ref{fig:norm_plot} shows that all the independent 
ways of estimating the scale factor are within 5\% of each other,
indicating robustness.

\subsection{How good is the isolated multiplet approximation for $\ell \leq 300$?} \label{sec:QDPT_vs_DPT}

\begin{figure}
    \centering
    \includegraphics[width=\textwidth]{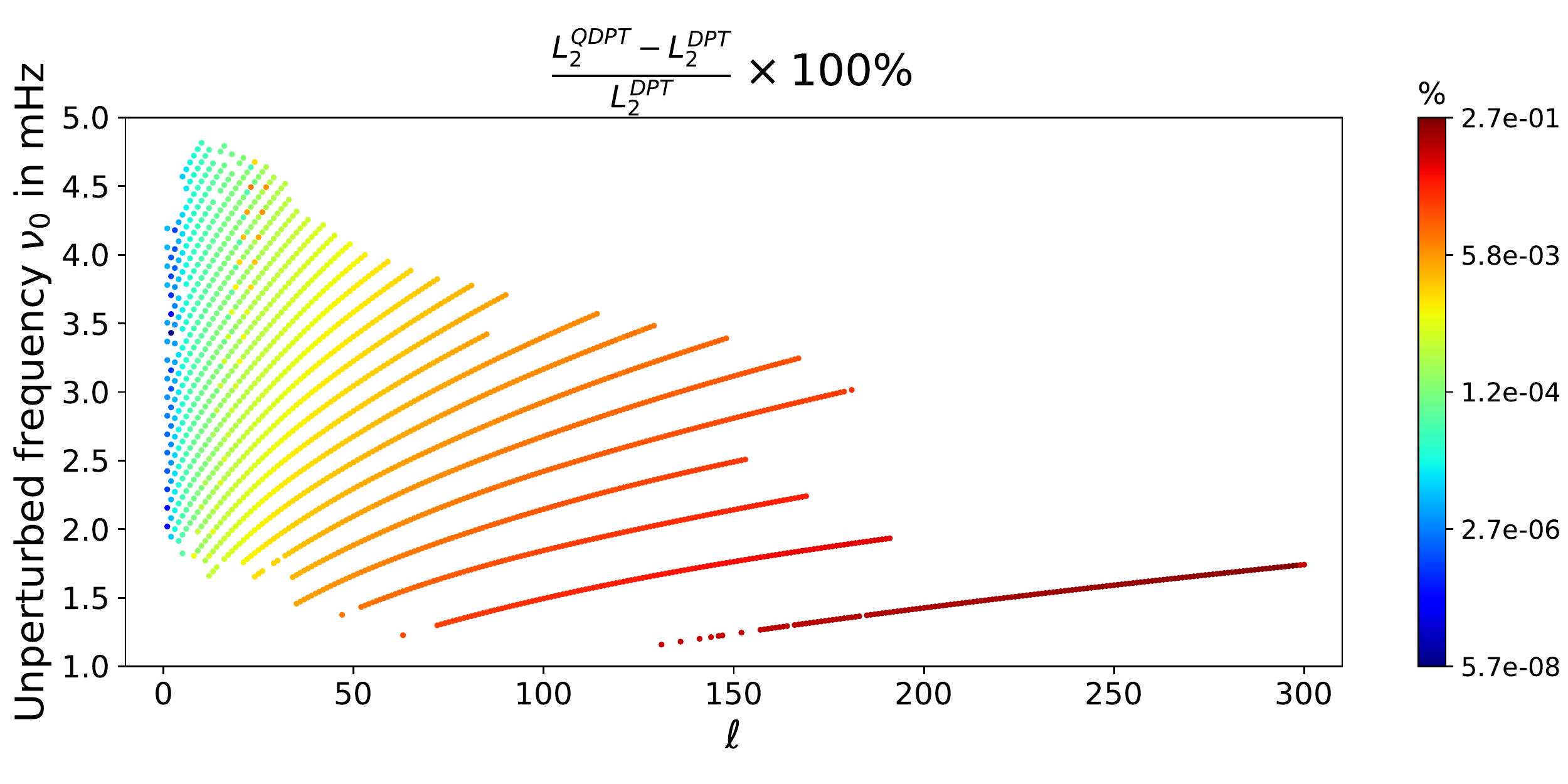}
    \caption{The relative offset of $L_2^\text{QDPT}$ as compared to that of $L_2^\text{DPT}$ (see Eqn~[\ref{eqn: l2_Q},\ref{eqn: l2_D}]) under the perturbation of an axisymmetric differential rotation $\Omega(r,\theta)$ as observed in the Sun. An increase in intensity of the color scale indicates worsening of the isolated multiplet approximation. The measures of offset are plotted for the HMI-resolved multiplets shown in Figure~\ref{fig:mode-selection}.}
    \label{fig:qdpt_vs_dpt_DR}
\end{figure}

Estimation of $a$ coefficients through frequency splitting measurements assumes validity of the isolated multiplet approximation using degenerate perturbation theory (DPT). However, an inspection of the distribution of the multiplets in $\nu-\ell$ space (as shown in Fig.~[\ref{fig:mode-selection}]) shows that it is natural to expect this approximation to worsen with increasing $\ell$. This necessitates carrying out frequency estimation respecting cross-coupling of modes across multiplets, also known as quasi-degenerate perturbation theory (QDPT).  A detailed discussion on DPT and QDPT in the context of differential rotation can be found in \cite{ritzwoller} and \cite{lavely92}. In this section we discuss the goodness of the isolated multiplet approximation in estimating $a_{n \ell}$ due to $\Omega(r,\theta)$ for all the HMI-resolved modes shown in Figure~\ref{fig:mode-selection}. A similar result but for $\ell \leq 30$ was presented in Appendix~G of \cite{sbdas20}. In Figure~\ref{fig:qdpt_vs_dpt_DR} we color code multiplets to indicate the departure of frequency shifts obtained from QDPT $\delta{}_{n}\omega{}_{\ell m}^{Q}$ as compared to shifts obtained from DPT $\delta{}_{n}\omega{}_{\ell m}^{D}$. Strictly speaking, carrying out the eigenvalue problem in the QDPT formalism causes garbling of the quantum numbers ---$n$, $\ell$, and~$m$ are no longer good quantum numbers--- and prevents a one-to-one mapping of unperturbed to perturbed modes. This prohibits an explicit comparison of frequency shifts on a singlet-by-singlet basis. However, modes belonging to the same multiplet can still be identified visually and grouped together. So, to quantify the departure of $\delta{}_{n}\omega{}_{\ell m}^{Q}$ from $\delta{}_{n}\omega{}_{\ell m}^{D}$ we calculate the Frobenius norm of these frequency shifts corresponding to each multiplet:
\begin{eqnarray}
    L_2^\text{QDPT} &=& \sqrt{\sum_m (\delta {}_{n}\omega{}_{\ell m}^\text{Q})^2} \qquad \text{for cross-coupling,} \label{eqn: l2_Q}\\
    L_2^\text{DPT} &=&  \sqrt{\sum_m (\delta {}_{n}\omega{}_{\ell m}^\text{D})^2} \qquad \text{for self-coupling.} \label{eqn: l2_D}
\end{eqnarray}
The color scale intensity in Figure~\ref{fig:qdpt_vs_dpt_DR} indicates the relative
offset of $L_2^\text{QDPT}$ as compared to $L_2^\text{DPT}$ for a multiplet 
$(n, \ell)$ marked as an `o'. Larger offset indicates the degree of worsening of 
the isolated multiplet approximation. We find that the largest error incurred
using DPT instead of QPDT is 0.27\% this is found to be at $\ell=300$. 
This clearly shows that even for the 
$f$ mode (which is the most susceptible to errors) the frequency splitting $a$-coefficients are exceptionally accurate.

\section{Conclusion} \label{sec:conclusion}

Most of what is currently known about solar differential rotation
is derived from from $a$-coefficients using frequency splitting measurements. 
Inferring these $a$-coefficients involves invoking the isolated multiplet
approximation based on degenerate perturbation theory. 
Although this approximation
works well even for high $\ell \leq 300$ modes, reasons 
motivating the need to investigate the possibility of erroneous 
$a$-coefficients from frequency splitting measurements at even higher
$\ell$ stem from a combination of two effects, namely, the increasing 
proximity of modes (in frequency) along the same radial branch, and spectral-leakage
from neighbouring modes.
Partial visibility of the Sun causes broadening of peaks in the 
spectral domain, referred to as mode leakage \citep{schou94, hanasoge18}. 
This causes proximal modes at high $\ell$ to widen and resemble 
continuous ridges in observed spectra.
As a result, spectral-peak identification for frequency-splitting
measurements are harder and increasingly inaccurate. Moreover, since 
the $a$-coefficient formalism breaks down for non-axisymmetric
perturbations, considering techniques which respect
cross-coupling becomes indispensable.
Thus, mode coupling becomes more relevant in these regimes, and it
is important to investigate the potential of mode-coupling techniques as
compared to frequency splittings. Hence, this study was directed towards
answering the following broad questions. (i) Can mode-coupling via MCMC
use information stored in eigenfunction distortions 
to constrain differential rotation as accurately as frequency splittings? This 
would also serve to compare the potential of a Bayesian approach with the 
least square inversion performed in W13.
(ii) Can this technique further increase the accuracy of $a_{n\ell}$ at 
$\ell \geq 150$?  We already know that higher $\ell$ estimates are increasingly
precise and accurate from W13. (iii) What are the uncertainties in estimating
$a_{n\ell}$ using mode-coupling theory and do they fall within 1-$\sigma$ 
of frequency splitting estimates?
(iv) Why are mode-coupling results poorer in the low $\ell$ regime? This is
seen in earlier studies, which aimed to go deeper into the convection zone 
and obtained significantly imprecise and inaccurate results \citep{woodard13,schad20}. 

The approach in this study is broadly based on the theoretical formulations
from V11 and modelling from W13. However, the novelty of 
the current work lies in 3 main aspects. (a) The MCMC analysis enabled 
exploration of the complete parameter space, and it was found that
the chosen misfit function is unimodal in nature, for all degrees~$\ell$
and radial orders~$n$. This establishes that 
the method of normal-mode coupling does return a unique value of 
$(a_3, a_5)$. 
(b) Leakage of power occurs for modes in the same radial order $n$ and 
hence the determination of $(a_3, a_5)$ in a consistent manner would
involve simultaneous estimation of splitting coefficients 
for all~$\ell$ and the same radial order. However, the number of parameters is
large and hence we break it into chunks of 40 pairs of 
$(a_3, a_5)$ per MCMC, with an overlap of $N_o$ pairs of the parameters
between two different chunks. To settle on a reasonable value of $N_o$, we
perform a simple experiment. From MCMC simulations with different overlap
numbers $N_o = \{0, 2, 4, 6, 8, 10\}$, we find that for $N_o > 6$, the inferred
$a$-coefficients vary less than 1-$\sigma$ and therefore reasonably stable for
larger $N_o$. Hence, we choose the modal overlap number $N_o=10$ for 
computation at all radial orders.
(c) Since a large number of splitting coefficients are determined
simultaneously, a corresponding number of spectra is used. Hence, 
estimation of the data variance becomes critical in order
to appropriately weight different data points according to their
noise levels. These improvements lead to a better estimate of differential
rotation using mode coupling.

The inference of rotation at lower $\ell$ ($<50$) suffers for
two reasons. (a) Low sensitivity of the model to the $a$-coefficients.
(b) Proximity of modes of radial orders $(n+1)$ and $(n-1)$ to modes at
radial order $n$. Since the current model only accounts for leakage of power
within the same radial order, a chosen frequency window in data would 
contain peaks from neighboring radial orders, which are not modelled. Hence,
an improvement might be achieved at lower $\ell$ by modelling the interaction
of modes of different radial orders. 

Finally, in this study we also show that even though
frequency splitting is much more precise for low $\ell \leq 150$, mode coupling 
estimates of differential rotation improves at high $\ell \geq 200$. Therefore, 
it is expected that mode-coupling would be comparable to 
(or possibly more accurate than) frequency splitting for very 
high $\ell \geq 300$. This would then
allow one to compare mode-coupling estimates of shallow, small-scale
structures with results from methods in local helioseismology. 
Going this high in angular degree for mode-coupling, however,
introduces some challenges: (a) The computation of
leakage matrices for high $\ell$ is very expensive. 
(b) $\partial \omega/\partial \ell$ decreases as $\ell$ grows 
and the spectrum becomes a continuous ridge in frequency space making
it harder to resolve the modes completely.

In conclusion, there remains scope for improvement and related lines 
of study. In this study, we have ignored the even-$s$ components of
$\Omega_s$, which are the NS-asymmetric components of differential rotation. 
These components have been estimated to be 
small at the surface and are anticipated to be small in the interior.
However, this assumption may be premature given that prior
estimates of interior rotation-asymmetries are based on non-seismic surface measurements. Since the V11 formalism 
is capable of accommodating the estimation of 
even-$s$ components as well, this could be the focus of a future investigation. 
Additionally, the current 
analysis was performed after summing up the stacked cross-spectrum.
Although this was done to improve the signal-to-noise ratio, 
the spectrum at different azimuthal orders $m$ are not identical. 
Hence a more complete computation would involve
the misfit computed using the full spectrum as a function of $m$. This may 
possibly lead to better results of the $a$-coefficients, as there exists
structure in the azimuthal order (see Fig.~[\ref{fig:cs-200-202}]), 
which is lost after summation. \\

The authors of this study are grateful to Jesper Schou (Max Planck Institute for Solar System Research) for numerous 
insightful discussions as well as detailed comments that helped us improve
the quality of the manuscript. The authors thank the anonymous referee
for valuable suggestions that helped improve the text and figures in this  manuscript.

\begin{appendices}

\section{Spherical harmonics symmetry relations}\label{apdx:sph-symm}
Consider a time-varying, real-valued scalar field on a sphere 
\(\phi(\theta, \phi, t)\). The
spherical harmonic components are given by
\begin{equation}
    \phi^{l,|m|}(t) = \int_\Omega \rmd\Omega Y^{*l, |m|}(\theta, \phi) 
    \phi(\theta, \phi, t)
    = (-1)^{|m|} \int_\Omega \rmd\Omega Y^{l, -|m|} \phi(\theta, \phi, t)
    = (-1)^{|m|} \phi^{*l, -|m|}(t)
\end{equation}
where $d\Omega$ is the area element, the integration being performed over
the entire surface of the sphere.
After performing a temporal Fourier transform, we have 
\begin{equation}
    \phi^{l,|m|}(\omega) = \frac{1}{\sqrt{2\pi}}\int_{-\infty}^\infty \rmd t 
    e^{-i\omega t} \phi^{l, |m|}(t) 
    = (-1)^{|m|} \frac{1}{\sqrt{2\pi}}\int_{-\infty}^\infty \rmd t 
    e^{-i\omega t} \phi^{* l, -|m|}(t) 
\end{equation}
\begin{equation}
\phi^{*l, -|m|}(\omega) = \frac{1}{\sqrt{2\pi}} \int_{-\infty}^\infty \rmd t e^{i\omega t}
\phi^{*l, -|m|}(t) = (-1)^{|m|} \phi^{l, |m|}(-\omega) \implies 
\phi^{l, -|m|}(\omega) = (-1)^{|m|} \phi^{*l, |m|}(-\omega)
\end{equation}

\section{MCMC: An illustrative Case} \label{sec:MCMC_demo}

\begin{figure}
    \centering
    \includegraphics[width=0.9\textwidth]
    {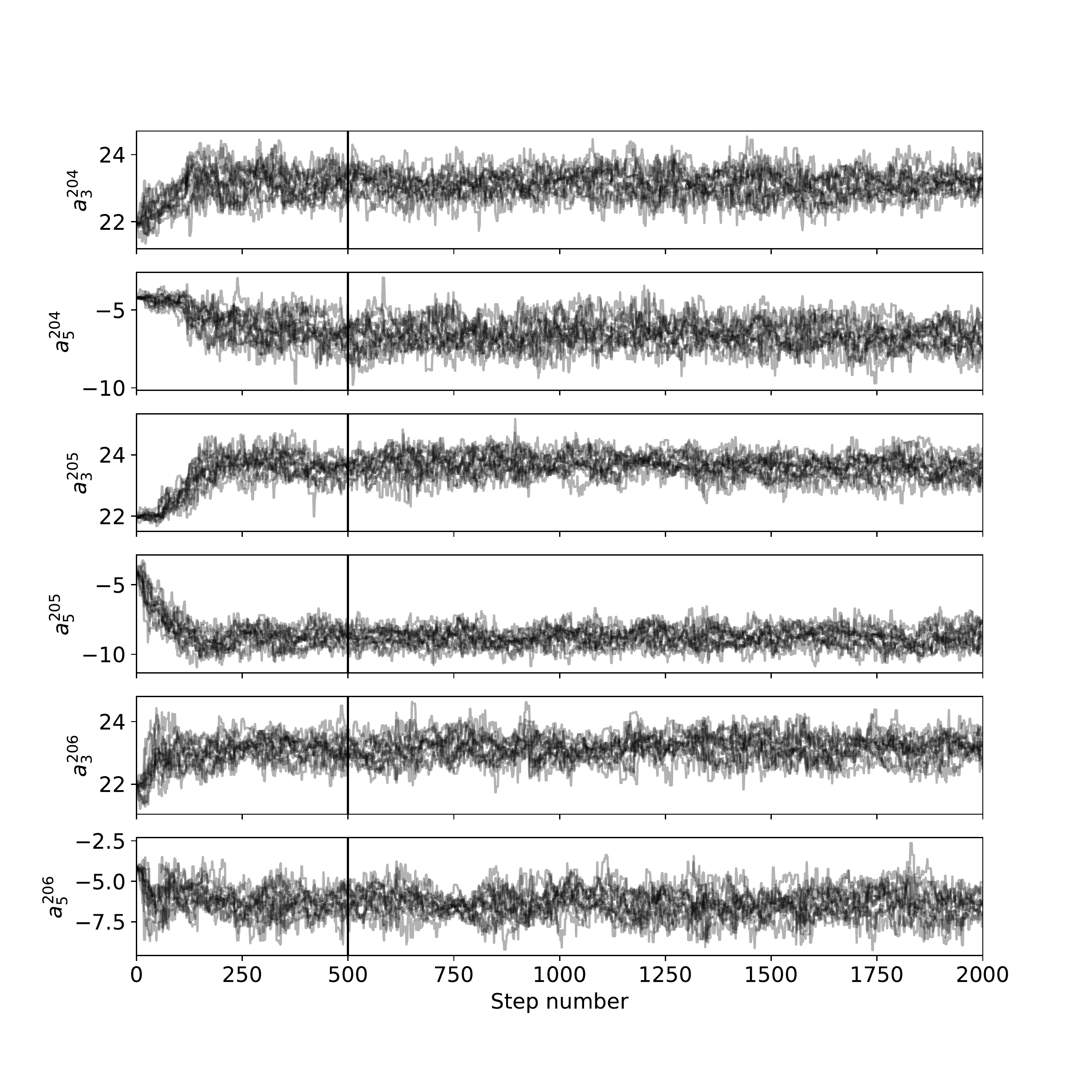}
    \caption{Each parameter is shown in a different figure to 
    indicate the value as a function of the Markov Chain step number.
    The first few ``burn-in" values are discarded and only the values
    beyond the vertical line are considered to obtain the probability
    distributions.}
    \label{fig:chain-sample}
\end{figure}

\begin{figure}
    \centering
    \includegraphics[width=0.9\textwidth]
    {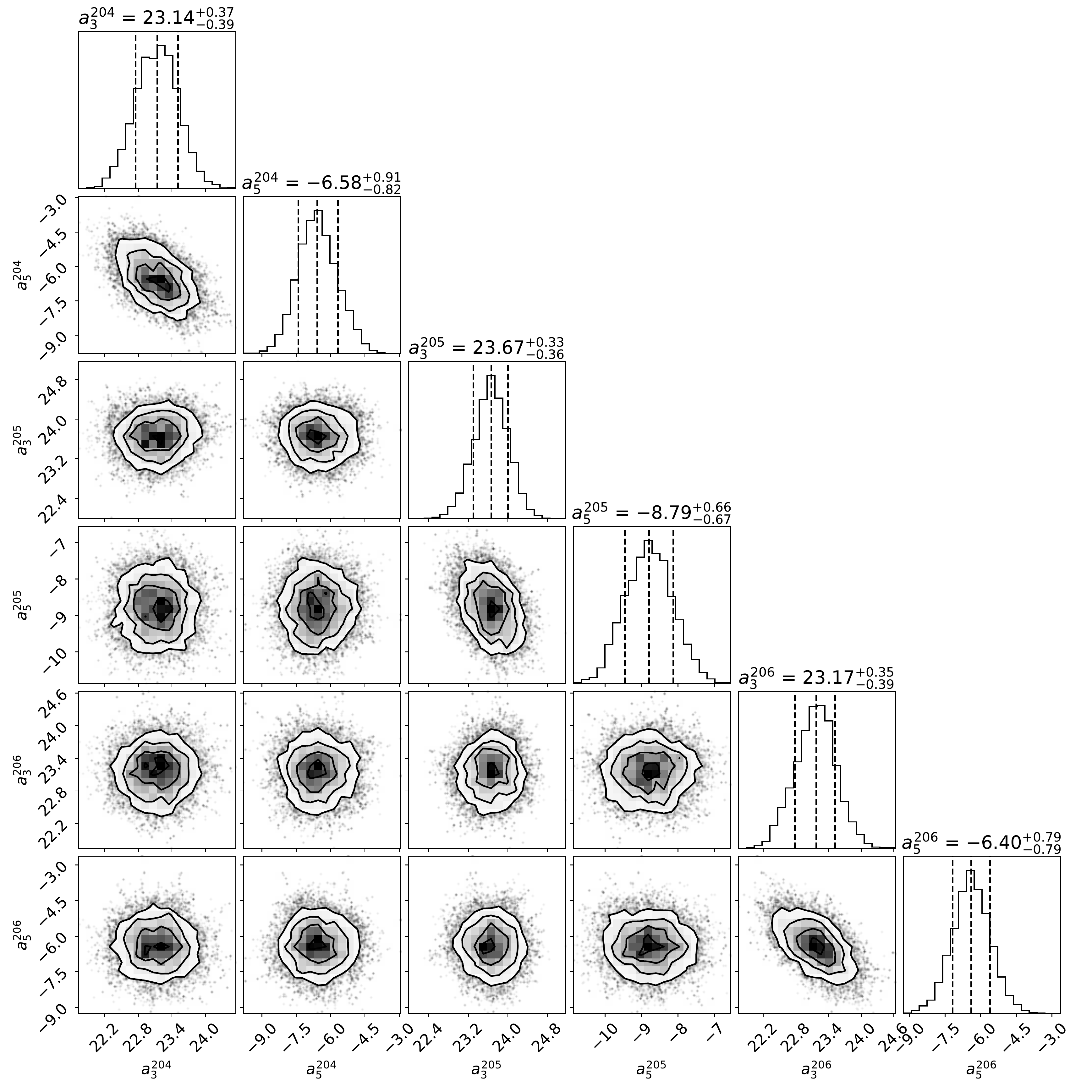}
    \caption{Cross-correlation of model parameters and the 
    marginal probability of the model parameters.}
    \label{fig:corner-sample}
\end{figure}

We present an MCMC estimation of $a$-coefficients using a smaller
set of modes (and hence model parameters). 
The smaller number of model parameters lets us present all the 
marginal probabilities in a single plot. The MCMC walkers 
are shown in Figure~\ref{fig:chain-sample}. 
In spite of using a flat prior, the likelihood function is sharp enough 
to bias the walkers to  move towards the region of optimal solution 
within $\sim 500$ iterations. 
It can be seen that different walkers start off randomly
at different locations in parameter space and ultimately converge
to the same region around the optimal solution. After removing 
the iterations from the ``burn-in" period, where the walkers are
still exploring a larger parameter space, histograms are plotted 
and marginal probability distributions are obtained.
Figure~\ref{fig:corner-sample} shows one such estimation of 
$(a_3, a_5)$ for $n=0$ and $\ell$ in the range $200$ to $202$. It can 
be seen that the marginal posterior probability distributions for
each of the parameters are unimodal. This tells us that the currently
defined misfit function has a unique minimum. Note that this 
distribution was obtained using a flat prior and hence the resulting
posterior distributions are essentially sampling the likelihood function.
It is also worth noting that for the range of $\ell$'s chosen, the
confidence intervals are $< 1$ nHz. 

\end{appendices}

\bibliography{references}{}
\bibliographystyle{aasjournal}





\end{document}

%% file: newcommands.tex
\newcommand{\bfxi}{\mbox{\boldmath $\bf \xi$}}

\def\rmd{{\mathrm{d}}}

\newcommand{\tdot}{\,.\hspace{-0.98 mm}\raise.6ex\hbox{.}
                   \hspace{-0.98 mm}\raise1.2ex\hbox{.}\,}

\def\bfv{{\mathbf{v}}}

\def\bfZ{{\mathbf{Z}}}

\renewcommand{\vec}[1]{\boldsymbol{#1}}